\documentclass[11pt]{article}
\usepackage{graphicx,verbatim,array,multicol,palatino,algorithm}
\usepackage{amsmath,amssymb,amsfonts,ifthen,setspace,color}
\newboolean{ShowFigures}
\newboolean{UnBlinded}
\newboolean{DoubleSpaced}
\setboolean{ShowFigures}{true}
\setboolean{UnBlinded}{true}
\setboolean{DoubleSpaced}{false}
\def\company#1{\textsf{\small #1}}

\def\sigeps{\sigma_{\varepsilon}}
\def\bib{\vskip11pt\par\noindent\hangindent=1 true cm\hangafter=1}
\def\bA{\boldsymbol{A}}
\def\bC{\boldsymbol{C}}
\def\bc{\boldsymbol{c}}
\def\be{\boldsymbol{e}}
\def\bB{\boldsymbol{B}}
\def\btheta{\boldsymbol{\theta}}
\def\bgamma{\boldsymbol{\gamma}}
\def\bb{\boldsymbol{b}}
\def\diag{\mbox{diag}}
\def\bOmega{\boldsymbol{\Omega}}
\def\bz{\boldsymbol{z}}
\def\bdeta{\boldsymbol{\eta}}
\def\diagonal{\mbox{diagonal}}
\def\bsigma{\boldsymbol{\sigma}}
\def\logit{\mbox{logit}}
\def\sigsqbeta{\sigma^2_{\beta}}
\def\qunder{\underline{q}}
\def\simind{\stackrel{{\tiny \mbox{ind.}}}{\sim}}
\def\xnew{x_{\mbox{{\tiny new}}}}
\def\ynew{y_{\mbox{{\tiny new}}}}
\def\bmu{\boldsymbol{\mu}}
\def\bSigma{\boldsymbol{\Sigma}}
\def\bZ{\boldsymbol{Z}}
\def\bbeta{\boldsymbol{\beta}}
\def\bX{\boldsymbol{X}}
\def\bu{\boldsymbol{u}}
\def\bv{\boldsymbol{v}}
\def\bw{\boldsymbol{w}}
\def\bx{\boldsymbol{x}}
\def\bxi{\boldsymbol{\xi}}
\def\by{\boldsymbol{y}}
\def\bI{\boldsymbol{I}}
\def\bzero{\boldsymbol{0}}
\def\bone{\boldsymbol{1}}
\def\punder{\underline{p}}
\def\smhalf{{\textstyle{\frac{1}{2}}}}
\def\thickboxit#1{\vbox{{\hrule height 1mm}\hbox{{\vrule width 1mm}\kern6pt
          \vbox{\kern6pt#1\kern6pt}\kern6pt{\vrule width 1mm}}
               {\hrule height 1mm}}}
\def\myand{\&\ }
\def\jump{\vskip3mm\noindent}
\def\fWO{f_{\mbox{\tiny WO}}}
\def\gammaDot{\gamma_{\bullet}}

\def\nwarm{n_{\mbox{\scriptsize warm}}}
\def\nvalid{n_{\mbox{\scriptsize valid}}}
\def\bywarm{\by_{\mbox{\scriptsize warm}}}
\def\bXwarm{\bX_{\mbox{\scriptsize warm}}}
\def\bCwarm{\bC_{\mbox{\scriptsize warm}}}
\def\bZwarm{\bZ_{\mbox{\scriptsize warm}}}
\def\bxiwarm{\bxi_{\mbox{\scriptsize warm}}}
\def\ynew{y_{\mbox{\tiny new}}}
\def\inew{i_{\mbox{\tiny new}}}
\def\snew{s_{\mbox{\tiny new}}}
\def\tnew{t_{\mbox{\tiny new}}}
\def\benew{\be_{\mbox{\tiny new}}}
\def\bxnew{\bx_{\mbox{\tiny new}}}
\def\bXnew{\bX_{\mbox{\tiny new}}}
\def\bcnew{\bc_{\mbox{\tiny new}}}
\def\bznew{\bz_{\mbox{\tiny new}}}
\def\aeps{a_{\varepsilon}}
\def\Aeps{A_{\varepsilon}}

\def\Au{A_{u}}
\def\auell{a_{u\ell}}
\begin{document}
\ifthenelse{\boolean{DoubleSpaced}}
{\setstretch{1.5}}{}

\begin{center}
{\LARGE\bf Real-time semiparametric regression}

\vskip4mm
{\sc By J. Luts$\null^1$, T. Broderick$\null^2$ and M.P. Wand$\null^1$}
\vskip4mm
{\it
$\null^1$
School of Mathematical Sciences,
University of Technology Sydney, Broadway 2007, Australia}
\vskip4mm
{\it 
$\null^2$
Department of Statistics,
University of California, Berkeley, California 94720, USA}
\end{center}

\vskip3mm
\centerline{4th February, 2013}
\vskip3mm
\centerline{\sc Summary}
\vskip2mm

We develop algorithms for performing semiparametric
regression analysis in real time, with data processed 
as it is collected and made immediately available via 
modern telecommunications technologies.
Our definition of semiparametric regression is quite broad and includes, 
as special cases, generalized linear mixed models, 
generalized additive models, geostatistical models, 
wavelet nonparametric regression models and their various combinations.
Fast updating of regression fits is achieved by couching 
semiparametric regression into a Bayesian hierarchical
model or, equivalently, graphical model framework
and employing online mean field variational ideas. An internet site
attached to this article,
\texttt{realtime-semiparametric-regression.net},
illustrates the methodology for continually 
arriving stock market, real estate and airline data. Flexible real-time 
analyses, based on increasingly ubiquitous streaming 
data sources stand to benefit.

\noindent
\jump

\section{Introduction}\label{sec:intro}

Ongoing technological advancements mean that data
are being collected and made available for inference
with rapidly increasing volume and speed. There
are numerous examples of this explosion of data,
but two that have established connections with 
semiparametric regression, our focus in this
article, are Internet auction analysis 
(e.g. Jank \myand Shmueli, 2007) and real-time 
spatial epidemiology (e.g. Kaimi \myand Diggle, 2011). 

\emph{Semiparametric regression} refers to a large
class of regression models that provide for non-linear
predictor effects using spline and wavelet basis
functions, as well as dependencies arising in grouped
data such as within-subject correlation. An arsenal
of both frequentist and Bayesian fitting and inference
procedures now exist. Recent overviews are contained
in Ruppert, Wand \myand Carroll (2009) and Wand \myand Ormerod (2011).

Virtually all semiparametric regression methodology 
proposed to date assume that the data are 
processed in \emph{batch}; that is, all at the same time.
Summaries such as function estimates, 
confidence intervals and posterior density functions 
are then outputted. Downsides to batch
processing include the requirement that statistical analysis
wait until an entire data set has been assembled
and, sometimes, the necessity of storing the
entire data set in 
memory. In the \emph{online} case, the procedure updates
as each new data point (or subset of data points) is obtained. Online
updates use only the new data and summary statistics from previous
iterations rather than the full set of available data. A particular
advantage of online processing is
that summaries, such as those just mentioned,
are updated throughout the data collection
process and therefore are available
immediately upon demand.
Online processing also has the
advantage of not requiring storage of potentially 
very large data-sets.

While a number of batch procedures exist for performing
semiparametric regression, we focus on a particular
methodology here due to the ease of adapting it to the online
framework as well as its wide range of applicability.
Consider single predictor nonparametric regression,
a special case of semiparametric regression with
a long history and large literature. Fully automatic
nonparametric regression batch procedures include:
(a) local linear kernel smoother
with cross-validation bandwidth selection,
(b) local linear kernel smoother with 
direct plug-in bandwidth selection,
(c) frequentist low-rank smoothing spline with
restricted maximum likelihood smoothing parameter selection,
(d) Bayesian low-rank smoothing spline with
Markov chain Monte Carlo approximate inference and
(e) Bayesian low-rank smoothing spline with
mean field variational Bayesian (MFVB) approximate inference.
Details of (a) are in H\"ardle (1990), details of 
(b) are in Wand \myand Jones (1995), whilst (c) and (d) 
are described in Ruppert, Wand \myand Carroll (2003).
Section 2.7 of Wand \myand Ormerod (2011) explains (e). 
Approaches (a)--(d) are more established, but none have a 
viable online modification. However, (e) is
relatively easy to modify for this purpose.

Another advantage of the Bayesian low-rank smoothing spline approach
to nonparametric regression is its extendibility. As explained
in Wand (2009), couching semiparametric regression in a graphical
models framework permits arbitrarily sophisticated models to 
be handled elegantly, efficiently, and cohesively.
This approach can handle
generalized 
additive models, geostatistical models,
wave\-let nonparametric 
regression models and their various combinations, as well
complications such as outliers and missingness.
Inference in these models is often accomplished by applying Markov
chain Monte Carlo procedures using the directed acyclic graph of 
variable dependencies. While versatile and accurate, such inference procedures
can be unacceptably slow. MFVB approaches, as demonstrated in Faes, 
Ormerod \myand Wand (2011) and Wand \myand Ormerod (2011), 
are a much faster alternative. Some accuracy and versatility 
must be sacrificed in return for the increased speed of MFVB. 
Nonetheless, for the models treated in this article MFVB accuracy ranges
from good to excellent.

Iterative algorithms that make a single pass through the
data -- with one iteration per
data point or per some small, fixed number of data points -- have
recently been developed for variational Bayesian inference.
In the machine learning literature,
Hoffman, Blei \myand Bach (2010) introduced such
an MFVB algorithm for latent Dirichlet allocation and
applied their algorithm to topic modeling. This
procedure was extended to the hierarchical 
Dirichlet process by Wang, Paisley \myand Blei (2011).
Tchumtchoua, Dunson \myand Morris (2012) further
developed online MFVB approximate inference for
high-dimensional correlated data. 
The methodology in these articles is 
referred to as \emph{online mean field variational Bayes}
or often with the shorter name \emph{online variational Bayes}.
While they are indeed single-pass and require storing at most a 
small, fixed number of data points in memory, 
they do, however, require knowledge of the number of
data points from the start of the algorithm.
Our focus in this work, by contrast, is not on transforming MFVB algorithms 
that require multiple data passes into single-pass algorithms.
Rather, we are, in some sense, pursuing a more classical definition of
an ``online algorithm'' in that each iteration of our procedure uses
past data only in the form of sufficient statistics and 
future data not at all.

Online MFVB has not been entertained previously
for nonparametric and semiparametric regression, but there is an
old and large literature involving other online approaches.
For nonparametric regression and the related 
density estimation problem Wolverton \myand Wagner (1969), 
Yamato (1971), Devroye \myand Wagner (1980) and 
Krzyak \myand Pawlak (1984) are examples of 
early articles on online analysis using 
kernel estimators.  However, they are chiefly
concerned with theoretical properties of the estimators
and are devoid of practical automatic smoothing 
parameter selection strategies.

Outside of semiparametric regression there are also
large literatures on online analysis. A few recent examples
are: Ng, McLachlan \myand Lee (2006) on 
prediction of hospital resource utilization,
Fricker \myand Chang (2008) on biosurveillance
and Kaimi \myand Diggle (2011) on monitoring of 
variation in risk of infections. A very recent article
by Michalak {\it et al.} (2012) describes the development 
of systems for real-time streaming analysis.

Semiparametric regression is a highly visual branch
of Statistics, with graphics being a crucial means
of conveying and diagnosing regression fits.
The norm for such graphical display are ink drawings
on pieces of paper or figures in PDF file.
Real-time semiparametric regression represents
a paradigm shift in graphical display, where 
regression summaries are best thought of as dynamic
graphics on web-pages or iDevice apps. We have
organized an Internet site that illustrates
real-time semiparametric regression graphical display.

Section \ref{sec:GaussResp} introduces the notion
of real-time semiparametric regression with
online MFVB via increasingly more sophisticated
Gaussian response models. Both classical and 
sparse shrinkage are treated. The more challenging 
binary response case is dealt with in Section \ref{sec:binResp}.
In Section \ref{sec:justif} we justify our approach
to real-time semiparametric regression in relation
to various other online learning methods such as
stochastic gradient descent. Some discussion about
inferential accuracy is given in Section \ref{sec:infAccur}.
Dynamic web-pages that illustrate the new methodology 
on live data are the focus of Section \ref{sec:illus}.

\section{Gaussian Response Models}\label{sec:GaussResp}

The conversion of a batch MFVB semiparametric
regression procedure to one that does online 
processing is particularly straightforward
in the Gaussian response case. We start
by explaining such conversion for the multiple
linear regression model, since it has 
minimal notational overhead.

\subsection{Multiple Linear Regression}\label{sec:mlr}

Let $\bX$ be a $n\times p$ design matrix and 
consider the Bayesian regression model
\begin{equation}
\by|\,\bbeta,\sigma^2\sim N(\bX\bbeta,\sigma^2\,\bI),\quad
\bbeta\sim N(\bzero,\sigma_{\beta}^2\,\bI),\quad
\sigma\sim\mbox{Half-Cauchy}(A).
\label{eq:linRegMod}
\end{equation}
where the $\mbox{Half-Cauchy}(A)$ prior is such that
the prior density function of $\sigma$ satisfies 
$p(\sigma)\propto\{1+(\sigma/A)^2\}^{-1},\ \sigma>0$.
An equivalent, but more tractable model, is that where
$\sigma\sim\mbox{Half-Cauchy}(A)$ 
is replaced by the auxiliary variable 
representation
\begin{equation}
\sigma^2|\,a\sim\mbox{Inverse-Gamma}(\smhalf,1/a),\quad
a\sim\mbox{Inverse-Gamma}(\smhalf,1/A^2)
\label{eq:HCtoIG}
\end{equation}
where the random variable $v\sim\mbox{Inverse-Gamma}(A,B)$
if and only if its density function is
$$p(v)=B^A\Gamma(A)^{-1}\,v^{-A-1}\,\exp(-v/B),\quad v>0.$$
A pertinent result for this distribution is $E(1/v)=A/B$.
Figure \ref{fig:linDAG} displays the directed acyclic graph
corresponding to the model conveyed by (\ref{eq:linRegMod}) 
and (\ref{eq:HCtoIG}).

\ifthenelse{\boolean{ShowFigures}}
{
\begin{figure}[!ht]
\centering
{\includegraphics[width=0.5\textwidth]{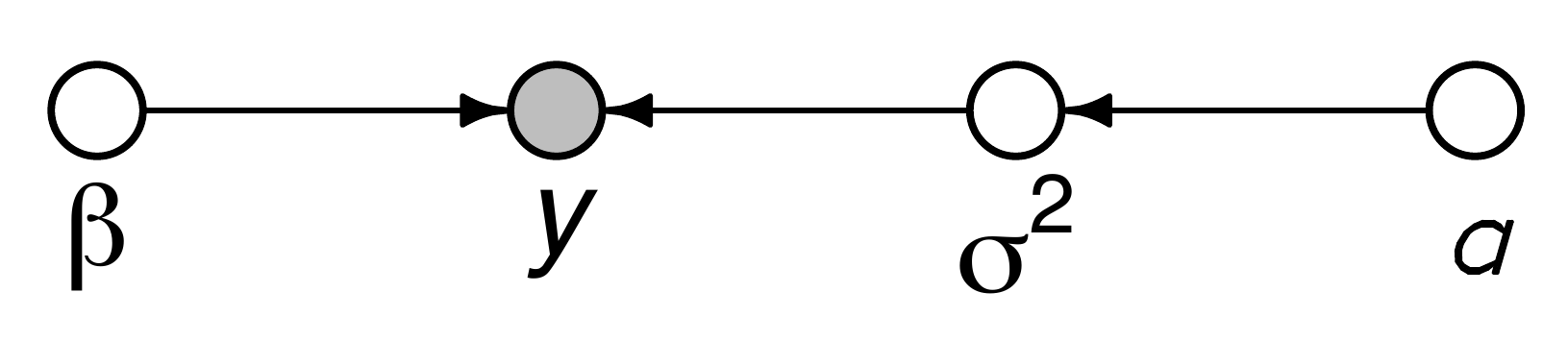}}
\caption{\it Directed acyclic graph for the 
model conveyed by (\ref{eq:linRegMod}) and (\ref{eq:HCtoIG}). 
The shading corresponds to the observed data.}
\label{fig:linDAG} 
\end{figure}
}
{\vskip3mm
\thickboxit{\bf \centerline{DAG figure here.}}
\vskip3mm
}

MFVB is a general prescription for approximation
of posterior density functions in a graphical model.
General references on MFVB include Bishop (2006) and
Wainwright \myand Jordan (2008).
Mean field approximation of the joint 
posterior density function $p(\bbeta,a,\sigma^2|\by)$ 
is founded upon this function being restricted to have 
a product form such as
\begin{equation}
q(\bbeta,a)\,q(\sigma^2)
\label{eq:prodRestrict}
\end{equation}
for density functions $q(\bbeta,a)$ and $q(\sigma^2)$.
We then choose these so-called $q$-density functions
to minimize the Kullback-Leibler distance between
$p(\bbeta,a,\sigma^2|\by)$ and $q(\bbeta,a)\,q(\sigma^2)$:
$$\int q(\bbeta,a)\,q(\sigma^2)\,
\log\left\{\frac{q(\bbeta,a)\,q(\sigma^2)}
{p(\bbeta,a,\sigma^2|\by)}\right\}\,d\bbeta\,da\,d\sigma^2.$$
Standard manipulations show that an equivalent optimization
problem is that of maximizing
$$\punder(\by;q)\equiv\exp\int 
q(\bbeta,a)\,q(\sigma^2)
\log\left\{\frac{p(\bbeta,a,\sigma^2,\by)}
{q(\bbeta,a)\,q(\sigma^2)}\right\}\,d\bbeta\,da\,d\sigma^2$$
and that $\punder(\by;q)$ is a lower bound on the 
marginal likelihood $p(\by)$ for all $q$-densities.
The solutions can be shown to satisfy 
\begin{equation}
{\setlength\arraycolsep{3pt}
\begin{array}{rcl}
q^*(\bbeta,a)&\propto&\exp[E_{q(\sigma^2)}\{\log\{p(\bbeta,a|\by,\sigma^2)\}],
\\[1ex]
\mbox{and}\ \ 
q^*(\sigma^2)&\propto&\exp[E_{q(\bbeta,a)}\{\log\{p(\sigma^2|\by,\bbeta,a)\}]
\end{array}
}
\label{eq:MFVBsolns}
\end{equation}
(see, e.g., Section 2.2 of Ormerod \myand Wand, 2010). 
Application of standard distribution theory to (\ref{eq:MFVBsolns})
shows that
\begin{equation}
\begin{array}{l}
q^*(\bbeta,a)\ \mbox{is the product of the}\ 
N(\bmu_{q(\bbeta)},\bSigma_{q(\bbeta)})\ \mbox{density function}\\[0.75ex]
\mbox{and the Inverse-Gamma}(1,B_{q(a)})\ \mbox{density function;}\\[0.75ex]
q^*(\sigma^2)\ \mbox{is the}\ 
\mbox{Inverse-Gamma}(\smhalf(n+1),B_{q(\sigma^2)})\ 
\mbox{density function}
\end{array}
\label{eq:qStarForms}
\end{equation}
for parameters $\bmu_{q(\bbeta)}$ and $\bSigma_{q(\bbeta)}$,
the mean vector and covariance matrix of $q^*(\bbeta)$, 
$B_{q(a)}$, the rate parameter of $q^*(a)$
and $B_{q(\sigma^2)}$, the rate parameter of $q^*(\sigma^2)$.
The MFVB solution is also such that 
$q^*(\bbeta,a)=q^*(\bbeta)\,q^*(a)$ even though 
(\ref{eq:prodRestrict}) does not assume this.

The symbols $\bmu_{q(\bbeta)}$ and $\bSigma_{q(\bbeta)}$
in (\ref{eq:qStarForms}) are instances of the following
general notation that we use throughout this article. 
If $v$ is a random variable having density function 
$q(v)$ then
$$\mu_{q(v)}\equiv E_q(v)\quad\mbox{and}\quad
\sigma^2_{q(v)}\equiv \mbox{Var}_q(v).
$$
If $\bv$ is a random vector having density function 
$q(\bv)$ then
$$\bmu_{q(\bv)}\equiv E_q(\bv)\quad\mbox{and}\quad
\bSigma_{q(\bv)}\equiv \mbox{Cov}_q(\bv).
$$

The optimal parameters in the $q^*$-density functions are 
interrelated. For example, 
$$\bSigma_{q(\bbeta)}=\left\{
\mu_{q(1/\sigma^2)}\,\bX^T\bX+\sigma_{\beta}^{-2}\,\bI
\right\}^{-1}.
$$
Hence, they must be 
obtained via an iterative coordinate ascent algorithm, 
in which equalities between the parameters are replaced by updates. 
This leads to Algorithm \ref{alg:lmBatchMFVB} for batch
MFVB fitting of (\ref{eq:linRegMod}) and (\ref{eq:HCtoIG}).
Each update is guaranteed to increase the value of 
$\punder(\by;q)$ (e.g. Luenberger \myand Ye, 2008).

\begin{algorithm}
\begin{center}
\begin{minipage}[t]{150mm}
\hrule
\begin{itemize}
\item[] Initialize:\ $\mu_{q(1/\sigma^2)}>0$.
\item[] Read in $\by\ (n\times 1)$ and $\bX\ (n\times p)$.  
\item[] Cycle:
\begin{itemize}
\item[] $\bSigma_{q(\bbeta)}\leftarrow
\left\{\mu_{q(1/\sigma^2)}\,\bX^T\bX+\sigma_{\beta}^{-2}\,\bI
\right\}^{-1}$
\item[] $\bmu_{q(\bbeta)}\leftarrow 
\mu_{q(1/\sigma^2)}\,\bSigma_{q(\bbeta)}\,\bX^T\by$\ \ \ \ ;\ \ \ \  
$\mu_{q(1/a)}\leftarrow 1/\{\mu_{q(1/\sigma^2)}+A^{-2}\}$
\item[]
$\mu_{q(1/\sigma^2)}\leftarrow 
\displaystyle{\frac{n+1}{2\,\mu_{q(1/a)}
+\by^T\by-2\bmu_{q(\bbeta)}^T\bX^T\by
+\mbox{tr}[(\bX^T\bX)\{\bSigma_{q(\bbeta)}+\bmu_{q(\bbeta)}
\bmu_{q(\bbeta)}^T\}]}}$
\end{itemize}
\item[] until the increase in $\punder(\by;q)$ is negligible.
\item[]
Produce summaries based on $q^*(\bbeta)\sim 
N(\bmu_{q(\bbeta)},\bSigma_{q(\bbeta)})$
and 
\item[]$q^*(\sigma^2)\sim
\mbox{Inverse-Gamma}(\smhalf(n+1),(n+1)/\{2\mu_{q(1/\sigma^2)}\})$.
\end{itemize}
\hrule
\end{minipage}
\end{center}
\caption{\it Batch mean field variational Bayes algorithm for
approximate inference in the Gaussian response linear regression
model (\ref{eq:linRegMod}) and (\ref{eq:HCtoIG}).}
\label{alg:lmBatchMFVB} 
\end{algorithm}

The lower bound on the marginal log-likelihood, used to monitor
convergence in Algorithm \ref{alg:lmBatchMFVB}, has
explicit expression
\begin{eqnarray*}
\log\,\punder(\by;q)&=&\smhalf\,p-\smhalf\,n\log(2\pi)-2\log(\pi)
+\log\Gamma(\smhalf(n+1))\\
&&\quad-\smhalf\,p\,\log(\sigma^2_{\bbeta})-\log(A)
-\textstyle{\frac{1}{2\sigma_{\bbeta}^2}}\{\Vert\bmu_{q(\bbeta)}\Vert^2
+\mbox{tr}(\bSigma_{q(\bbeta)})\}
\\
&&\quad
+\smhalf\log|\bSigma_{q(\bbeta)}|
-\smhalf(n+1)\log[(n+1)/\{2\mu_{q(1/\sigma^2)}\}]
\\
&&\quad
-\log(\mu_{q(1/\sigma^2)}+A^{-2})
+\mu_{q(1/\sigma^2)}\mu_{q(1/a)}.
\end{eqnarray*}

In Algorithm \ref{alg:lmBatchMFVB},
dependence on the data is only through the quantities
$\by^T\by$, $\bX^T\by$ and $\bX^T\bX$ and each of
these have simple updates when a new response $\ynew$
and its corresponding $p\times 1$ vector of predictors
$\bxnew$ arrives. For example, the new $\bX^T\bX$ matrix
is 
$$\bXnew^T\bXnew=\bX^T\bX+\bxnew\bxnew^T.$$
Based on these observations Algorithm \ref{alg:lmOnlineMFVB},
the \emph{online} modification of the Algorithm \ref{alg:lmBatchMFVB},
ensues.

\begin{algorithm}
\begin{center}
\begin{minipage}[t]{150mm}
\hrule
\begin{itemize}
\item[] Initialize:\ $\mu_{q(1/\sigma^2)}>0$, 
\ $\by^T\by\leftarrow 0$,
\ $\bX^T\,\by\leftarrow\bzero_{p\times 1}$,
\ $\bX^T\bX\leftarrow\bzero_{p\times p}$,
\ $n\leftarrow 0$.
\item[] Cycle:
\begin{itemize}
\item[] read in $\ynew\ (1\times 1)$ and $\bxnew\ (p\times 1)$
\ \ ;\ \ $n\leftarrow n+1$
\item[] $\by^T\by\leftarrow\by^T\by+\ynew^2$\ \ ;\ \ 
$\bX^T\by\leftarrow\bX^T\by+\bxnew\,\ynew$\ \ ;\ \
$\bX^T\bX\leftarrow\bX^T\bX+\bxnew\,\bxnew^T$
\item[] $\bSigma_{q(\bbeta)}\leftarrow
\left\{\mu_{q(1/\sigma^2)}\,\bX^T\bX+\sigma_{\beta}^{-2}\,\bI
\right\}^{-1}$
\item[] $\bmu_{q(\bbeta)}\leftarrow 
\mu_{q(1/\sigma^2)}\,\bSigma_{q(\bbeta)}\,\bX^T\by$\ \ ;\ \ 
$\mu_{q(1/a)}\leftarrow 1/\{\mu_{q(1/\sigma^2)}+A^{-2}\}$
\item[]
$\mu_{q(1/\sigma^2)}\leftarrow 
\displaystyle{\frac{n+1}{2\,\mu_{q(1/a)}
+\by^T\by-2\bmu_{q(\bbeta)}^T\bX^T\by
+\mbox{tr}[(\bX^T\bX)\{\bSigma_{q(\bbeta)}+\bmu_{q(\bbeta)}
\bmu_{q(\bbeta)}^T\}]}}$
\item[]produce summaries based on $q^*(\bbeta)\sim 
N(\bmu_{q(\bbeta)},\bSigma_{q(\bbeta)})$
and 
\item[]$q^*(\sigma^2)\sim
\mbox{Inverse-Gamma}(\smhalf(n+1),(n+1)/\{2\mu_{q(1/\sigma^2)}\})$
\end{itemize}
\item[] until data no longer available or analysis terminated.
\end{itemize}
\hrule
\end{minipage}
\end{center}
\caption{\it Online mean field variational Bayes algorithm for
approximate inference in the Gaussian response linear regression
model (\ref{eq:linRegMod}) and (\ref{eq:HCtoIG}).}
\label{alg:lmOnlineMFVB} 
\end{algorithm}

Algorithm \ref{alg:lmOnlineMFVB} differs from Algorithm
\ref{alg:lmBatchMFVB} in that the data are processed 
on arrival and the approximate posterior densities
of the model parameters are continually updated.
In the case of streaming data there is the option of 
dynamic graphical displays of the approximate posterior density 
functions of the regression coefficients and error variance 
and corresponding approximate Bayes estimates and credible sets.
Dynamic regression diagnostic plots could also be entertained.

\ifthenelse{\boolean{ShowFigures}}
{
\begin{figure}[!ht]
\centering
{\includegraphics[width=1.0\textwidth]{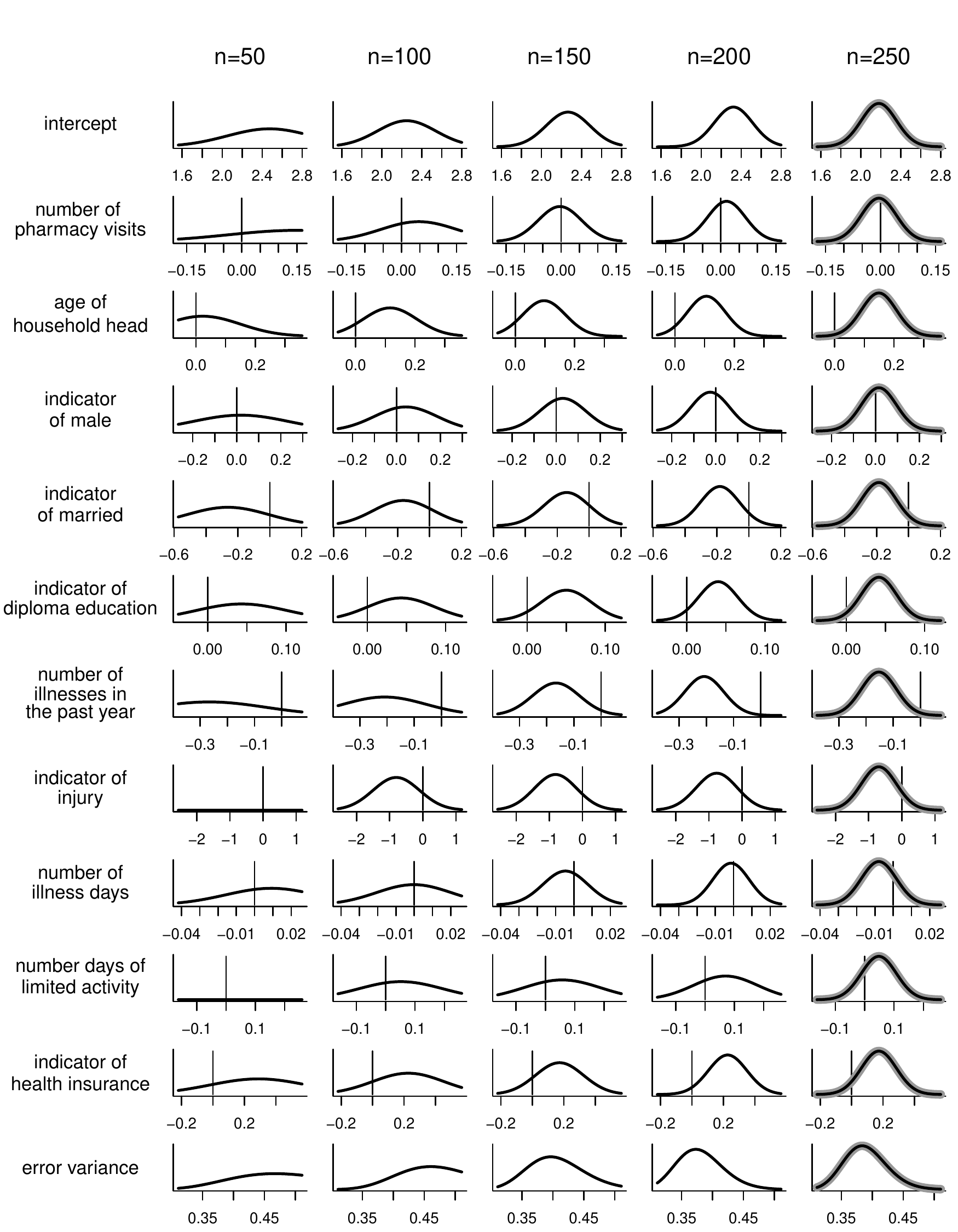}}
\caption{\it Successive approximate posterior density functions of regression 
coefficients and the logarithm of error variance for the 
Vietnam medical expenses data described in the text.
The predictors corresponding to each regression coefficient
are listed in the left-hand columns.
The posterior density functions are based on online MFVB 
as detailed in Algorithm \ref{alg:lmOnlineMFVB}. 
The axis limits are the same across each
row and a vertical line is positioned at zero.  
For $n=250$ the batch MFVB approximate fits are shown
as thick grey curves.
}
\label{fig:vietnamMFVB} 
\end{figure}
}
{\vskip3mm
\thickboxit{\bf \centerline{Vietnam figure here.}}
\vskip3mm
}

Figure \ref{fig:vietnamMFVB} provides rudimentary illustration 
of online regression inference when data from the
Vietnam World Bank Living Standards Survey 
(source: Cameron \myand Trivedi, 2005)
are fed into Algorithm \ref{alg:lmOnlineMFVB}.
These data are in the \texttt{VietNamI} data-frame of the 
\textsf{R} package \texttt{Ecdat} (Croissant, 2011).
The response variable is the logarithm of total medical
expenses.  Description of the predictor variables is given
in the \texttt{VietNamI} documentation of Crossaint (2011).
Each variable was transformed to lie inside 
the unit interval before being processed. The scaling is 
determined using an initialization batch just as for the 
initial parameter tuning described in Section \ref{sec:tuning}.
The posterior density functions were then 
back-transformed to correspond to the original units.
The  hyperparameters were set at $\sigma^2_{\bbeta}=10^{10}$ 
and $A=10^5$ to impose non-informativity.

Note, for example, the approximate posterior density functions
for $\beta_2$, the regression coefficient attached to 
\texttt{age of household head}. For $n\le100$ the posterior density 
function is relatively flat and $\beta_2$ is not statistically
significant. As $n$ increases, the posterior density functions
become narrower and, by $n=250$, the lower limit of
the 95\% credible set is positive -- indicating statistical
significance of this predictor. 

The right-most column of Figure \ref{fig:vietnamMFVB}
shows the batch MFVB posterior density functions for $n=250$.
In this case, the batch and online MFVB results are seen
to be virtually identical. However, as demonstrated later,
such agreement is not guaranteed in general.

\subsubsection{Batch-based Tuning and Convergence Diagnosis}
\label{sec:tuning}

\ifthenelse{\boolean{ShowFigures}}
{
\begin{figure}[!ht]
\centering
{\includegraphics[width=\textwidth]{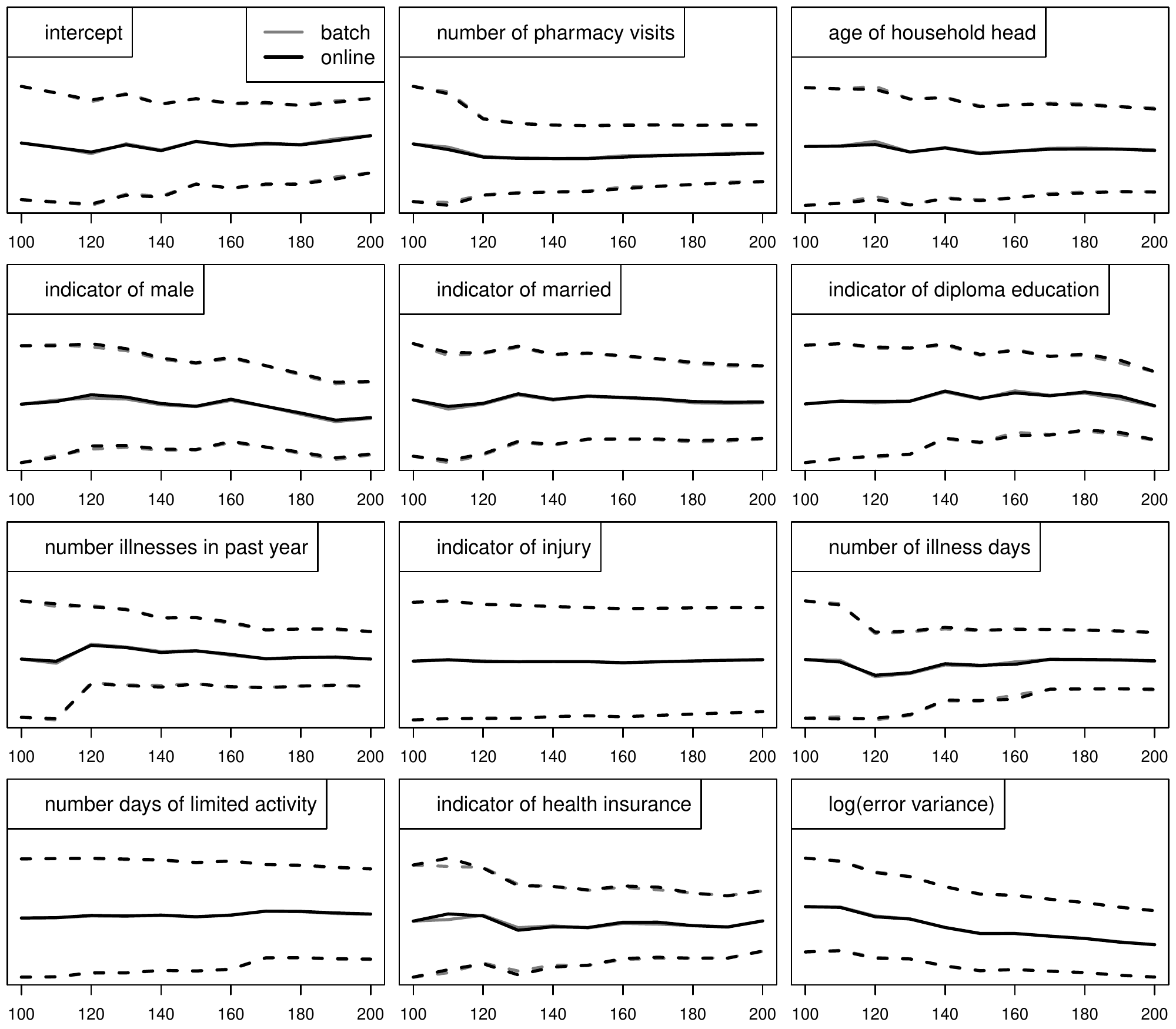}}
\caption{\it Convergence diagnostics for the example
given in Figure \ref{fig:vietnamMFVB}. 
The solid lines track the posterior means, whilst
the dashed lines show corresponding 95\% credible sets.
The horizontal axes show the sample sizes between a warm-up batch
sample of size $\nwarm=100$ and validation sample sizes up to 
$\nvalid=100$ greater than $\nwarm$.}
\label{fig:vietnamConvDiag} 
\end{figure}
}
{\vskip3mm
\thickboxit{\bf \centerline{Vietnam diagnostic figure here.}}
\vskip3mm
}

Ideally, the online Algorithm \ref{alg:lmOnlineMFVB}
will mimic the results of the batch Algorithm \ref{alg:lmBatchMFVB}
as the sample size $n$ increases. However, we know of no
guarantees that this will happen and it is possible
that the online parameters will diverge from their
batch counterparts. For the more elaborate models studied
later in this article, such divergence is very common.
Therefore, convergence diagnosis at the start of the 
online iterations is essential. The principal idea
is to start by running a small subset of initial data points in
the batch algorithm to obtain starting values for both
data sufficient statistics and, more importantly, estimated
parameters of the model. A second, small validation subset of data
is used to compare the batch and online algorithm results. If convergence of the
online iterations to their batch counterparts
is not verified by this comparison then larger initial
batch runs are required to tune the online algorithm.

The idea of collecting streaming data into a small subset before
processing it in order to improve performance of
a single-pass algorithm is reminiscent of the
``mini-batches'' of Hoffman, Blei \myand Bach (2010).
However, in our approach, the batching of data happens
only with a small subset at the very beginning of the algorithm rather
than throughout. Also, we develop an alternative tuning method for
this subset batch size below; notably, our tuning method requires batching
only some initial subset of the data rather than the full data set.

We will now provide
details via the Figure \ref{fig:vietnamMFVB} example.
Figure \ref{fig:vietnamConvDiag} shows the posterior
means and 95\% credible sets for each $\beta_j$, $0\le j\le 11$,
and $\log(\sigma^2)$ and sample sizes $n=100,110,\ldots,200$
when the Vietnam medical expenses data are fitted via both
batch and online MFVB. The batch MFVB summary statistics
(shown as grey lines in Figure \ref{fig:vietnamConvDiag})
correspond to simply inputting the first $\nwarm=100$
observations into Algorithm \ref{alg:lmBatchMFVB} and
then repeating this process for 10 additional equally-spaced sample
sizes that are $\nvalid=100$ greater than $\nwarm$. The
largest sample size is then $\nwarm+\nvalid=200$.
The online results (shown as grey lines in Figure \ref{fig:vietnamConvDiag})
were obtained via online MFVB updating steps of Algorithm \ref{alg:lmOnlineMFVB} 
but with $\mu_{q(1/\sigma^2)}$, $\by^T\by$, 
$\bX^T\by$, $\bX^T\bX$ and $n$ initialized at
\emph{the values obtained when the first $\nwarm=100$} observations
are inputted into Algorithm \ref{alg:lmBatchMFVB}.
This implies that all the results are identical at $n=\nwarm=100$,
but there are some small discrepancies for $n>100$. 
In this example the discrepancies are negligible, and hard to discern
from Figure \ref{fig:vietnamConvDiag} -- indicating convergence
of the online MFVB algorithm.  Figure \ref{fig:binRespConvDiag} in 
Section \ref{sec:binResp} shows an example where convergence
is not achieved with $\nwarm=100$ and a larger warm-up is required.

Algorithm 2' is a modification of Algorithm \ref{alg:lmOnlineMFVB} 
that incorporates batch-based tuning and convergence 
diagnostics. Whilst such modification is not necessary for the 
example depicted in Figures \ref{fig:vietnamMFVB} and 
\ref{fig:vietnamConvDiag}, it is crucial for more sophisticated
semiparametric models such as those described later 
in this article.

\begin{center}
\begin{minipage}[t]{150mm}
\hrule
\begin{itemize}
\item[]
\item[1.] Set $\nwarm$ to be the warm-up sample size
and $\nvalid$ to be size of the validation period.
Read in the first $\nwarm+\nvalid$ response and predictor
values.
\item[2.] Create $\bywarm$ and $\bXwarm$ consisting of the
first $\nwarm$ response and predictor values.
\item[3.] Feed $\bywarm$ and $\bXwarm$ into the
batch MFVB Algorithm \ref{alg:lmBatchMFVB} to obtain 
a starting value for $\mu_{q(1/\sigma^2)}$.
\item[4.] Set $\by^T\by\leftarrow \bywarm^T\bywarm$,
\ $\bX^T\,\by\leftarrow \bXwarm^T\bywarm$,
\ $\bX^T\bX\leftarrow \bXwarm^T\bXwarm$\\ and
$n\leftarrow \nwarm$.
\item[5.] Run the online MFVB Algorithm \ref{alg:lmOnlineMFVB}
until $n=\nwarm+\nvalid$.
\item[6.] Use convergence diagnostic graphics to assess whether
the online parameters are converging to the batch parameters.
\begin{itemize}
\item[(a)] If not converging then return to Step 1 and 
increase $\nwarm$. 
\item[(b)] If converging then continue running the online 
MFVB Algorithm \ref{alg:lmOnlineMFVB} until data no
longer available or analysis terminated.
\end{itemize}
\end{itemize}
\hrule
\end{minipage}
\end{center}
\jump
Algorithm 2': {\it Modification of Algorithm \ref{alg:lmOnlineMFVB}
to include batch-based tuning and convergence diagnosis.}

One could contemplate automating Step 6 of 
Algorithm 2', to save the user from having to 
conduct diagnostic checks. However, we have not yet
explored automatic convergence diagnosis and, instead,
flag this as a problem worthy of future research.

\subsubsection{Model Assumptions}

The online MFVB Algorithm 2' is founded upon 
the same assumptions as its batch counterpart 
Algorithm \ref{alg:lmBatchMFVB}. Both algorithms fit 
the Bayesian linear regression
model (\ref{eq:linRegMod}), but the latter
has the option to do the fitting in real time
for sequentially arriving data. 

Throughout this article, we are not allowing for the
model parameters to change as new data arrive. Colloquially, 
we assume ``fixed targets'' rather than ``moving targets''.
Extensions to semiparametric regression scenarios where the 
model parameters drift over time, and real-time algorithms that 
adapt to such drifts, are certainly worthy of future investigation --
but beyond this article's scope. 

\subsection{Linear Mixed Models}\label{sec:LMM}

A very useful structure for semiparametric regression
is the class of Bayesian linear mixed models of the form
\begin{equation}
\begin{array}{c}
\by|\,\bbeta,\bu,\sigeps^2 \sim 
N(\bX\bbeta+\bZ\bu,\sigeps^2\,\bI)\\[3ex]
\bu|\,\sigma_{u1}^2,
\ldots,\sigma_{ur}^2\sim N(\bzero,\mbox{blockdiag}
(\sigma_{u1}^2\,\bI_{K_1},\ldots,\sigma_{ur}^2\,\bI_{K_r}))
\end{array}
\label{eq:LMM1}
\end{equation}
where $\by$ is an $n\times1$ vector of response variables, 
$\bbeta$ is a $p\times1$ vector of fixed effects, 
$\bu$ is a vector of random effects, $\bX$ and $\bZ$ 
corresponding design matrices, $\sigeps^2$ is the error
variance and $\sigma_{u1}^2,\ldots,\sigma_{ur}^2$
are variance parameters corresponding to sub-blocks
of $\bu$ of size $K_1,\ldots,K_r$.
Here the priors are taken to be
\begin{equation}
\bbeta\sim N(\bzero,\sigma_{\beta}^2\bI),\quad
\sigma_{u\ell}\sim\mbox{Half-Cauchy}(A_{u\ell}),\ 1\le\ell\le r,\quad 
\sigeps\sim\mbox{Half-Cauchy}(A_{\varepsilon})
\label{eq:LMM2}
\end{equation}
with the hyperparameters satisfying
$\sigma_{\beta}^2,A_{\varepsilon},A_{u\ell}>0$
for $1\le\ell\le r$. As in Section \ref{sec:GaussResp}, 
tractability considerations motivate the introduction
of auxiliary variables 
\begin{equation}
a_{u\ell}\sim\mbox{Inverse-Gamma}(\smhalf,1/A_{u\ell}^2)
\quad\mbox{and}\quad
a_{\varepsilon}\sim\mbox{Inverse-Gamma}(\smhalf,1/A_{\varepsilon}^2)
\label{eq:LMM3}
\end{equation}
and use of the analogue of (\ref{eq:HCtoIG}) to induce Half-Cauchy
priors on the standard deviation parameters.

As spelt out in Section 2 of Zhao, Staudenmayer, Coull \myand Wand (2006),
model (\ref{eq:LMM1})--(\ref{eq:LMM2}) encompasses
a rich class of models including (with example number from 
Zhao {\it et al.}\ 2006 added):
\begin{itemize}
\item simple random effects models (Examples 1 and 2),
\item cross random effects models (Example 3),
\item nested random effects models (Example 4),
\item generalized additive models (Example 6),
\item semiparametric mixed models (Example 7),
\item bivariate smoothing and geoadditive models 
extensions (Example 8).
\end{itemize}
Examples 2 and 6 of Zhao {\it et al.}\ (2006) 
actually involve $2\times2$ and
$3\times3$ unstructured covariance matrix
parameters which are not covered by (\ref{eq:LMM2}).
However, as discussed in Section \ref{sec:unstrucCov},
the unstructured covariance matrix extension is quite straightforward.

We seek a mean field approximation to the joint
posterior density function: 
$$p(\bbeta,\bu,a_{u1},\ldots,a_{ur},a_{\varepsilon},
\sigma_{u1}^2,\ldots,\sigma_{ur}^2,\sigeps^2|\by)
\approx
q(\bbeta,\bu,a_{u1},\ldots,a_{ur},a_{\varepsilon},
\sigma_{u1}^2,\ldots,\sigma_{ur}^2,\sigeps^2).$$
The product form
\begin{equation}
q(\bbeta,\bu,a_{u1},\ldots,a_{ur},a_{\varepsilon},
\sigma_{u1}^2,\ldots,\sigma_{ur}^2,\sigeps^2)=
q(\bbeta,\bu,a_{u1},\ldots,a_{ur},a_{\varepsilon})\,   
q(\sigma_{u1}^2,\ldots,\sigma_{ur}^2,\sigeps^2).
\label{eq:prodResLMM}
\end{equation}
has the advantage of being minimally restrictive
whilst also yielding closed form MFVB updates.
The analogue of (\ref{eq:MFVBsolns}) leads to
$$
\begin{array}{l}
q^*(\bbeta,\bu,a_{u1},\ldots,a_{ur},a_{\varepsilon})
\ \mbox{is the product of the}\ 
N(\bmu_{q(\bbeta,\bu)},\bSigma_{q(\bbeta,\bu)})\ \mbox{density function,}\\[0.75ex]
\mbox{Inverse-Gamma}(1,B_{q(a_{u\ell})})\ 
\mbox{density functions, $1\le\ell\le r$,\ and the}\\[0.75ex]
\mbox{Inverse-Gamma}(1,B_{q(a_{\varepsilon})})\mbox{\ density function;}\\[1ex]
q(\sigma_{u1}^2,\ldots,\sigma_{ur}^2,\sigeps^2)\ \mbox{is the product of}
\mbox{\ Inverse-Gamma}(\smhalf(K_{\ell}+1),B_{q(\sigma^2_{u\ell})})
\mbox{\ density functions}\\[0.75ex]
\mbox{for $1\le\ell\le r$ and
the Inverse-Gamma}(\smhalf(n+1),B_{q(\sigeps^2)})\ \mbox{density function.}
\end{array}
$$
The subscripted $B$s are rate parameters. 
Batch MFVB fitting of (\ref{eq:LMM1}), but with 
slightly different prior distributions, is
given by Algorithm 3 of Ormerod \myand Wand (2010), where
the notation
$$\bC=[\bX\ \bZ]$$
is used. Let $P$ be the number of columns in $\bC$. Then each pass of
the corresponding online MFVB algorithm involves arrival and processing
of a new scalar response measurement, $\ynew$, and a
$P\times 1$ vector $\bcnew$, corresponding to the new row of $\bC$.
This results in Algorithm \ref{alg:lmmOnlineMFVB} for real-time
fitting of (\ref{eq:LMM1}).

\begin{algorithm}
\begin{center}
\begin{minipage}[t]{151mm}
\hrule
\begin{itemize}
\item[1.] Perform batch-based tuning runs analogous
to those described in Algorithm 2'
and determine a warm-up sample size $\nwarm$ for which
convergence is validated. 
\item[2.] Set $\bywarm$ and $\bCwarm$ to be the response vector
and design matrix based on the first $\nwarm$ observations.
Then set
\ $\by^T\by\leftarrow \bywarm^T\bywarm$,\ 
\ $\bC^T\,\by\leftarrow \bCwarm^T\bywarm$,
\ $\bC^T\bC\leftarrow\bCwarm^T\bCwarm$,
\ $n\leftarrow\nwarm$. Also, set
$\mu_{q(1/\sigeps^2)}$ and 
$\mu_{q(1/\sigma_{u1}^2)},\ldots,
\mu_{q(1/\sigma_{ur}^2)}$
to be the values for these 
quantities obtained in the batch-based tuning run
with sample size $\nwarm$.
\item[3.] Cycle:
\begin{itemize}
\item[] read in $\ynew\ (1\times 1)$ and $\bcnew\ (P\times 1)$
\ \ ;\ \ $n\leftarrow n+1$
\item[] $\by^T\by\leftarrow\by^T\by+\ynew^2$\ \ ;\ \ 
$\bC^T\by\leftarrow\bC^T\by+\bcnew\,\ynew$\ \ ;\ \
$\bC^T\bC\leftarrow\bC^T\bC+\bcnew\,\bcnew^T$
\item[] $\bSigma_{q(\bbeta,\bu)}\leftarrow
\left[\mu_{q(1/\sigeps^2)}\,\bC^T\bC+
\mbox{blockdiag}\{\sigma_{\beta}^{-2}\,\bI_p,
\mu_{q(1/\sigma_{u1}^2)}\bI_{K_1},\ldots,
\mu_{q(1/\sigma_{ur}^2)}\bI_{K_r}\}
\right]^{-1}$
\item[] $\bmu_{q(\bbeta,\bu)}\leftarrow 
\mu_{q(1/\sigeps^2)}\,\bSigma_{q(\bbeta,\bu)}\,\bC^T\by$\ \ \ ;\ \ \  
$\mu_{q(1/\aeps)}\leftarrow 1/\{\mu_{q(1/\sigeps^2)}+\Aeps^{-2}\}$
\item[]
$\mu_{q(1/\sigeps^2)}\leftarrow 
\displaystyle{\frac{n+1}{2\,\mu_{q(1/\aeps)}
+\by^T\by-2\bmu_{q(\bbeta,\bu)}^T\bC^T\by
+\mbox{tr}[(\bC^T\bC)\{\bSigma_{q(\bbeta,\bu)}+\bmu_{q(\bbeta,\bu)}
\bmu_{q(\bbeta,\bu)}^T\}]}}$
\item[] For $\ell=1,\ldots,r$\,:
\begin{itemize}
\item[]
$\mu_{q(1/a_{u\ell})}\leftarrow 
1/\{\mu_{q(1/\sigma_{u\ell}^2)}+A_{u\ell}^{-2}\}$
\item[]
$\mu_{q(1/\sigma_{u\ell}^2)}\leftarrow 
\displaystyle{\frac{K_{\ell}+1}{2\,\mu_{q(1/\auell)}
+\Vert\bmu_{q(\bu_{\ell})}\Vert^2
+\mbox{tr}(\bSigma_{q(\bu_{\ell})})}}$
\end{itemize}
\end{itemize}
\item[] until data no longer available or analysis terminated.
\end{itemize}
\hrule
\end{minipage}
\end{center}
\caption{\it Online mean field variational Bayes algorithm for
approximate inference in the Gaussian response linear mixed model
(\ref{eq:LMM1}).}
\label{alg:lmmOnlineMFVB} 
\end{algorithm}
 
The $\bcnew$ vector will have different
forms depending on the type of linear mixed model. To better 
understand the nature of these forms, consider the following
two special cases of (\ref{eq:LMM1}):
\begin{equation}
\begin{array}{c}
y_{ij}|\beta_0,U_i,\beta_1,\sigeps^2\simind
N(\beta_0+U_i+\beta_1\,x_{ij},\sigeps^2),\quad 1\le i\le m,
\quad 1\le j\le n_i,\\[1ex]
U_i|\,\sigma_u^2\simind N(0,\sigma_u^2\,\bI),\quad
\beta_0,\beta_1\simind N(0,\sigma_{\bbeta}^2),\\[1ex]
\sigma_u\sim\mbox{Half-Cauchy}(A_u),\quad
\sigeps\sim\mbox{Half-Cauchy}(A_{\varepsilon})
\end{array}
\label{eq:randInt}
\end{equation}
and 
\begin{equation}
\begin{array}{c}
y_i|\beta_0,\beta_s,\beta_t,\bu_s,\bu_t,\sigeps^2\simind
N\left(\beta_0+\beta_s\,s_i+\beta_t\,t_i+
{\displaystyle\sum_{k=1}^{K_s}}\,u_{s,k}\,z_k^s(s_i)+
{\displaystyle\sum_{k=1}^{K_t}}\,u_{t,k}\,z_k^t(t_i),
\sigeps^2\right),\\[1ex]
\quad 1\le i\le n,\quad \bu_s=[u_{s,1},\ldots,u_{s,K_s}]^T,
\quad \bu_t=[u_{t,1},\ldots,u_{t,K_t}]^T,\\[1ex]
\bu_s|\,\sigma_{u,s}^2\sim N(0,\sigma_{u,s}^2\,\bI),\quad
\bu_t|\,\sigma_{u,t}^2\sim N(0,\sigma_{u,t}^2\,\bI),\quad
\\[1ex]
\beta_0,\beta_s,\beta_t\simind N(0,\sigma_{\bbeta}^2),\quad
\sigma_{u,s}\sim\mbox{Half-Cauchy}(A_{u,s}),\quad
\sigma_{u,t}\sim\mbox{Half-Cauchy}(A_{u,t}),\\[1ex]
\sigeps\sim\mbox{Half-Cauchy}(A_{\varepsilon}).
\end{array}
\label{eq:bivAddModel}
\end{equation}
Here and throughout $\simind$ denotes ``distributed independently''.

Model (\ref{eq:randInt}) is the random intercept extension
of simple linear regression for longitudinal data with 
$(x_{ij},y_{ij})$ denoting the $j$th predictor/response pair 
for the $i$th group, with $m$ denoting the number of groups. 
There is no intrinsic reason to insist that the 
observations arrive in order with respect to the $i,j$
subscripting. Hence $\bcnew$ will have the form
$$\bcnew=\left[
\begin{array}{c}        
1 \\
\xnew \\
\benew  \\
\end{array}
\right]
$$
where $\xnew$ is the new predictor measurement that partners 
$\ynew$ and $\benew$ is a $m\times 1$ vector with an entry
of 1 in position $\inew$, corresponding to the group that
$(\xnew,\ynew)$ is from, and zeroes elsewhere.

Model (\ref{eq:bivAddModel}) is a mixed model-based
penalized spline version of the additive model
$$y_i=\beta_0+f_s(s_i)+f_t(t_i)+\varepsilon_i,\quad 1\le i\le n,$$
where the $s_i$ and $t_i$ are continuous predictor measurements
and $f_s$ and $f_t$ are smooth functions. The functions 
$z_k^s(\cdot)$, $1\le k\le K_s$, are spline basis functions.
A simple example is the truncated line basis 
\begin{equation}
z_k^s(s)=(s-\kappa^s_k)_+
\label{eq:truncLine}
\end{equation}
where $\kappa^s_1,\ldots,\kappa^s_{K_s}$ are a set of
knots within the domain of the $s_i$ values. More
sophisticated, and numerically stable, options for 
$z_k(s)$ are described in, for example, Wood (2006), 
Welham {\it et al.} (2007) and Wand \myand Ormerod (2008).
We use the last of these, known as O'Sullivan splines,
in our examples. The $z_k^t(\cdot)$, $1\le k\le K_t$, are 
defined similarly. A key feature of the $z_k^s(\cdot)$ and 
$z_k^t(\cdot)$ is that the multiple-of-diagonal covariance
matrices are appropriate under mixed model representations
of penalized splines. This subtlety is explained in 
Section 4 of Wand \myand Ormerod (2008).

Online fitting of (\ref{eq:bivAddModel})
involves reading in vectors of the form
$$\bcnew=[1,\snew,\tnew,z_1^s(\snew),\ldots,z_{K_s}^s(\snew),
z_1^t(\tnew),\ldots,z_{K_t}^t(\tnew)]^T
$$
where $\snew$ and $\tnew$ are the new predictor measurements
that partner $\ynew$. There is, however, the issue of having
to set the spline basis functions in advance. For instance, if 
the truncated line basis (\ref{eq:truncLine}) is used then
the knots have to be set at or near the start of the algorithm.
For many applications this is not a major problem. For example,
if the $\snew$ values correspond to age, in years, of human adults
then the range of possible $s_i$ values is easy to specify and
a reasonable spline basis can be set in advance. In a similar
vein, for longitudinal data, Algorithm \ref{alg:lmmOnlineMFVB} 
assumes that the number of groups is set in advance. If the
groups correspond to the counties of a geographical entity
then this should not pose a problem. If the data are from
a medical study then Algorithm \ref{alg:lmmOnlineMFVB}
assumes that the number of patients and their identity
numbers are fixed in advance. If this is not a reasonable
assumption then some adjustment is required. 

Finally, we mention the possibility of speeding up
the most expensive update:
\begin{equation}
\bSigma_{q(\bbeta,\bu)}\leftarrow
\left[\mu_{q(1/\sigeps^2)}\,\bC^T\bC+
\mbox{blockdiag}\{\sigma_{\beta}^{-2}\,\bI_p,
\mu_{q(1/\sigma_{u1}^2)}\bI_{K_1},\ldots,
\mu_{q(1/\sigma_{ur}^2)}\bI_{K_r}\}
\right]^{-1}.
\label{eq:CTCupdate}
\end{equation}
For Model (\ref{eq:randInt}) the matrix requiring inversion has 
dimension $(2+m)\times(2+m)$. If the number of groups is 
high then na\"{\i}ve implementation could lead to 
a bottleneck at (\ref{eq:CTCupdate}).
In the batch case it is well-known (e.g. Smith \myand Wand, 2008) 
that $\bC^T\bC$ contains diagonal forms that allow $O(m)$
computation of the right-hand side of (\ref{eq:CTCupdate}).
Such efficiencies are available in the online case, but
require careful  rearrangement of the entries of $\bC^T\bC$
during the updates.

\subsection{Extension to Unstructured Covariance Matrices for Random Effects}
\label{sec:unstrucCov}

A \emph{random intercepts and slopes} extension of (\ref{eq:randInt}) is
one with the first two hierarchical levels set to
$$
\begin{array}{c}
y_{ij}|\beta_0,\beta_1,U_i,V_i,\sigeps^2\simind
N(\beta_0+U_i+(\beta_1+V_i)\,x_{ij},\sigeps^2),\quad 1\le i\le m,
\quad 1\le j\le n_i,\\[2ex]
\mbox{and}\quad
\left[
\begin{array}{c}
U_i\\
V_i \\
\end{array}
\right]\Big|\bSigma\sim N(\bzero,\bSigma),
\quad\mbox{where}\quad
\bSigma\equiv  \left[\begin{array}{cc}         
\sigma_u^2 & \rho_{uv}\,\sigma_u\,\sigma_v \\
\rho_{uv}\,\sigma_u\,\sigma_v & \sigma_v^2
\end{array}\right]
\end{array}
$$
is an unstructured $2\times2$ covariance matrix.
The conjugate prior for $\bSigma$ is the Inverse Wishart distribution.
However, the specification
$$
\begin{array}{c}
\bSigma|\,a_{uv1},a_{uv2}\sim\mbox{Inverse-Wishart}\left(\nu+1,2\nu\,
\left[
\begin{array}{cc}
1/a_{uv1} & 0 \\
   0  & 1/a_{uv2} 
\end{array}
\right]
\right),\\[2ex]
a_{uv1},a_{uv2}\simind
\mbox{Inverse-Gamma}(\smhalf,1/A_{uv}),\quad \nu,A_{uv}>0
\end{array}
$$
provides a covariance matrix extension
of $\sigma_u\sim\mbox{Half-Cauchy}(A_u)$.
The choice $\nu=2$ is particularly attractive since
it imposes a $\mbox{Uniform}(-1,1)$ distribution on $\rho_{uv}$
and Half-$t_2$ distributions on $\sigma_u$ and $\sigma_v$.
This is laid out in Huang \myand Wand (2012), including
the definition of the $\mbox{Inverse-Wishart}(a,\bB)$ 
distribution.

Extensions to more sophisticated models, possibly having
larger unstructured covariance matrices, can be done in
a similar fashion.

\subsection{Extension to Sparse Shrinkage Penalties}\label{sec:sparse}

Model (\ref{eq:LMM1}) involves the following Gaussian
penalization on sub-vectors of $\bu$:
\begin{equation}
\bu_{\ell}|\,\sigma_{u\ell}^2\sim N(\bzero,\sigma_{u\ell}^2\,\bI),
\quad 1\le\ell\le r.
\label{eq:nonSparseu}
\end{equation}
However, many models of current-day interest, such 
as \emph{wide data (``$p\gg n$'')} and \emph{wavelet}
regression, require an assumption that the regression
coefficients are sparse. Under such sparseness assumptions,
the Gaussian priors (\ref{eq:nonSparseu}) are not appropriate
since they induce a relatively gentle amount of penalization
that lacks the ability to annihilate regression coefficients during
fitting and inference.

For simplicity of exposition we will confine discussion of the
sparse shrinkage extension to the $r=1$ version of (\ref{eq:LMM1}). 
Hence we retain
$$y\,|\,\bbeta,\bu,\sigeps^2\sim N(\bX\bbeta+\bZ\,\bu,\sigeps^2\bI)$$
without any sub-division of $\bu$. Let $K$ be the dimension of $\bu$
and consider general mutually independent prior penalizations
of the form:
$$u_k\simind p(u;\sigma_u,\btheta)$$
where $p(\cdot;\sigma_u,\btheta)$ is a density function
with scale parameter $\sigma_u$ and shape parameter $\btheta$.
Options for $p(u;1,\btheta)$ include:
\begin{equation}
{\setlength\arraycolsep{1pt}
\begin{array}{rcll}
p(u;1,w)&=&w\{\smhalf\exp(-|u|)\}+(1-w)\,\delta_0(u)&\  
\mbox{(Laplace-Zero),}  \\[2ex]
p(u;1)&=&(2\pi^3)^{-1/2}\exp(u^2/2)E_1(u^2/2)&\ 
\mbox{(Horseshoe),} \\[2ex]
p(u;1,\lambda)&=&\frac{\lambda\,2^{\lambda}\Gamma(\lambda+\smhalf)}{\pi^{1/2}}
\,\exp(u^2/4)
\,D_{-2\lambda-1}(|u|)
&\ \mbox{(Normal-Exponential-Gamma)} \\[2ex]
\mbox{and}\quad 
p(u;\lambda)&=&\displaystyle{\frac{1}{2(1+|u|/\lambda)^{\lambda+1}}}&\ 
\mbox{(Generalized Double Pareto).} \\
\end{array}
}
\label{eq:sparsePriors}
\end{equation}
Here $\delta_0$ denotes the Dirac delta function with 
mass at zero. Also, $E_1$ denotes the exponential integral function of order 1
and $D_{\nu}$ denotes the parabolic cylinder function of order $\nu$
according to the definitions of Gradshteyn \myand Ryzhik (1994).
References for the development of these sparse shrinkage priors
are Johnstone \myand Silverman (2005) (Laplace-Zero), Carvalho,
Polson \myand Scott (2010) (Horseshoe), Griffin \myand Brown (2011)
(Normal-Exponential-Gamma) and Armagan, Dunson \myand Lee (2012)
(Generalized Double Pareto).

Batch MFVB algorithms for the priors (\ref{eq:sparsePriors}) recently
have been derived by Wand \myand Ormerod (2011) 
(Laplace-Zero prior) and Neville, Ormerod \myand Wand (2012)
(Horseshoe, Normal-Exponential-Gamma and Generalized Double Pareto 
priors). 

Algorithm \ref{alg:sparseOnlMFVB} is the online adaptation of
Algorithm 4 of Wand \myand Ormerod (2011) for the Laplace-Zero 
prior model:
\begin{equation}
\begin{array}{c}
\by|\bbeta,\bgamma,\bv,\sigeps^2\sim 
N(\bX\bbeta+\bZ(\bgamma\odot\bv),\sigeps^2\bI),\quad
\bv|\,\sigma_u^2,\bb\sim N(\bzero,\sigma_u^2\,\diag(\bb)^{-1}),\\
\null\\
\sigma^2_u|\,a_u\sim\mbox{Inverse-Gamma}(\smhalf,1/a_u),\quad 
\sigeps^2|\,a_{\varepsilon}\sim\mbox{Inverse-Gamma}(\smhalf,1/a_{\varepsilon}),\\
\null\\
\bbeta\sim N(\bzero,\sigma_{\beta}^2\bI),
\ a_u\sim\mbox{Inverse-Gamma}(\smhalf,1/\Au^2),
\ a_{\varepsilon}\sim\mbox{Inverse-Gamma}(\smhalf,1/\Aeps^2),\\
\null\\
b_k\simind\mbox{Inverse-Gamma}(1,\smhalf),
\quad \gamma_k|\,\rho\sim\mbox{Bernoulli}(\rho),\quad 
\rho\simind\mbox{Beta}(A_{\rho},B_{\rho}). 
\end{array}
\label{eq:penWavBayeIG}
\end{equation}
Note that $\bA\odot\bB$ denotes the element-wise product
of matrices $\bA$ and $\bB$ having the same dimensions.
Model (\ref{eq:penWavBayeIG}) is a reproduction of (30) in
Wand \myand Ormerod (2011) and the additional notation is
explained there. Note, in particular, that the Laplace-Zero
prior is handled via the introduction of auxiliary variables
$\bb$, $\bgamma$ and $\bv$. Section 3.6 of Wand \myand Ormerod (2011)
provides the necessary details.
Similar online MFVB algorithms for the continuous sparse signal 
shrinkage priors listed in (\ref{eq:sparsePriors}) follow from
the batch MFVB algorithms of Neville, Ormerod \myand Wand (2012).

\begin{algorithm}
\begin{center}
\begin{minipage}[t]{150mm}
\hrule
\vskip2mm\noindent
\begin{itemize}
\item[1.] Perform batch-based tuning runs analogous
to those described in Algorithm 2'
and determine a warm-up sample size $\nwarm$ for which
convergence is validated. The batch
MFVB algorithm is Algorithm 4 of Wand \myand Ormerod (2011).
\item[2.] Set $\bywarm$ and $\bCwarm=[\bone\ \bZwarm]$ 
to be the response vector
and design matrix based on the first $\nwarm$ observations. 
Then set
\ $\by^T\by\leftarrow\bywarm^T\bywarm$,
\ $\bZ^T\bone\leftarrow\bZwarm^T\bone$,
\ $\bZ^T\by\leftarrow\bZwarm^T\by$,
\ $\bZ^T\bZ\leftarrow\bZwarm^T\bZwarm$,
\ $\bC^T\,\by\leftarrow \bCwarm^T\bywarm$,
\ $\bC^T\bC\leftarrow\bCwarm^T\bCwarm$,
\ $n\leftarrow\nwarm$. 
Set $K$ to be the number of columns in $\bZwarm$.
Also, set
$\mu_{q(1/\sigma^2_{\varepsilon})},\ 
\mu_{q(1/\sigma^2_{u})},\ \mu_{q(1/a_{\varepsilon})},\ 
\mu_{q(1/a_u)}, \bmu_{q(\bb)}, \bmu_{q(\bw_{\bgamma})}\ \mbox{and}
\ \bOmega_{q(\bw_{\bgamma})}$
to be the values for these
quantities obtained in the batch-based tuning run
with sample size $\nwarm$.
\item[3.] Cycle:
\begin{itemize}
\item[] read in $\ynew\ (1\times 1)$
and $\bznew\ (K\times 1)$
\ \ ;\ \ $n\leftarrow n+1$\ \ ;\ \ 
$\bcnew\leftarrow \left[\begin{array}{c}1\\ \bznew\end{array}\right]$
\item[] $\by^T\by\leftarrow\by^T\by+\ynew^2$\ \ ;\ \ 
$\bZ^T\bone\leftarrow\bZ^T\bone+\bznew$\ \ ;\ \
$\bZ^T\by\leftarrow\bZ^T\by+\bznew\,\ynew$
\item[]
$\bZ^T\bZ\leftarrow\bZ^T\bZ+\bznew\,\bznew^T$\ \ ;\ \
$\bC^T\by\leftarrow\bC^T\by+\bcnew\,\ynew$\ \ ;\ \
$\bC^T\bC\leftarrow\bC^T\bC+\bcnew\,\bcnew^T$
\item[] 
$\bSigma_{q(\bbeta,\bv)}\leftarrow\Bigg(\mu_{q(1/\sigeps^2)}
(\bC^T\bC)\odot\bOmega_{q(\bw_{\bgamma})}+\left[\begin{array}{cc}
\sigma_{\beta}^{-2} & \bzero \\
\bzero     & \mu_{q(1/\sigma_u^2)}\diag(\bmu_{q(\bb)})
\end{array}
\right]\Bigg)^{-1}
$
\item[] $\bmu_{q(\bbeta,\bv)}\leftarrow \mu_{q(1/\sigeps^2)}
           \bSigma_{q(\bbeta,\bv)}
           \diag\{\bmu_{q(\bw_{\bgamma})}\}\bC^T\by$ 
\item[] $\mu_{q(\bb)}\leftarrow
[\mu_{q(1/\sigma_u^2)}\{\mbox{diagonal}(\Sigma_{q(\bv)})
+\bmu^2_{q(\bv)}\}]^{-1/2}$
\vspace{-6mm}
\item[] 
\begin{eqnarray*}
\bdeta_{q(\bgamma)}&\leftarrow&-\smhalf\,\mu_{q(1/\sigeps^2)}\Big[
\diagonal(\bZ^T\bZ)\odot\{\bsigma^2_{q(\bv)}+\bmu^2_{q(\bv)}\}
-2(\bZ^T\by)\odot\bmu_{q(\bv)}\\
&&\quad+2(\bZ^T\bone)\odot\{[\bSigma_{q(\bbeta,\bv)})]_{i=1,2\le j\le K+1}      
+\mu_{q(\beta)}\bmu_{q(\bv)}\}\\
&&\quad +2\,\diagonal\{\bZ^T\bZ\,\diag(\bmu_{q(\bgamma)})\bSigma_{q(\bv)}\}\\
&&\quad
-2\,\diagonal(\bZ^T\bZ)\odot\bmu_{q(\bgamma)}\odot\diagonal(\bSigma_{q(\bv)})\\
&&\quad+2\bmu_{q(\bv)}\odot
\{\bZ^T\bZ(\bmu_{q(\bgamma)}\odot\bmu_{q(\bv)})
-\diagonal(\bZ^T\bZ)\odot\bmu_{q(\bgamma)}\odot\bmu_{q(\bv)}\}
\Big]\\
&&\quad+\psi(A_{\rho}+\mu_{q(\gammaDot)})-\psi(B_{\rho}+K-\mu_{q(\gammaDot)})
\end{eqnarray*}
\item[] $\mu_{q(\bgamma)}\leftarrow \displaystyle
{\frac{\exp(\eta_{q(\bgamma)})}{1+ \exp(\eta_{q(\bgamma)})}}$
\ \ \ ;\ \ \ 
$\bmu_{q(\bw_{\bgamma})}\leftarrow\left[
\begin{array}{c}  
1 \\
\bmu_{q(\bgamma)}
\end{array}
  \right]$\ \ \ ;\ \ \ 
$\mu_{q(\gammaDot)}\leftarrow\sum_{k=1}^K\mu_{q(\gamma_k)}$
\item[]
$\bOmega_{q(\bw_{\bgamma})}\leftarrow
\diag\{\bmu_{q(\bw_{\bgamma})}\odot
(\bone-\bmu_{q(\bw_{\bgamma})})\}+
\bmu_{q(\bw_{\bgamma})}\,\bmu_{q(\bw_{\bgamma})}^T
$
\item[] $\mu_{q(1/a_{\varepsilon})}\leftarrow 
1/\{\mu_{q(1/\sigeps^2)}+A_{\varepsilon}^{-2}\}$ \ \ ;\ \ 
$\mu_{q(1/a_u)}\leftarrow 
1/\{\mu_{q(1/\sigma_u^2)}+A_{u}^{-2}\}$
\item[] \hspace{-3mm}
$
\begin{array}{l}
B_{q(\sigeps^2)}\leftarrow
\mu_{q(1/a_{\varepsilon})}+\smhalf\,\by^T\by
-
\left(\bmu_{q(\bw_{\bgamma})}\odot\mu_{q(\bbeta,\bv)}\right)^T
\,\bC^T\by\\
\qquad\qquad+\smhalf\mbox{tr}\left(\bC^T\bC\,\,\left[
\bOmega_{q(\bw_{\bgamma})}
\odot\left\{\bSigma_{q(\bbeta,\bv)}+\bmu_{q(\bbeta,\bv)}
\bmu_{q(\bbeta,\bv)}^T
\right\}\right]\right)
\end{array}
$
\item[] $B_{q(\sigma^2_{u})}\leftarrow 
       \mu_{q(1/a_u)}+\smhalf\bmu_{q(\bb)}^T
        \{\mbox{diagonal}(\bSigma_{q(\bv)})+\bmu_{q(\bv)}^2\}$ 
\item[] $\mu_{q(1/\sigma_u^2)}\leftarrow 
\smhalf(K+1)/B_{q(\sigma^2_{u})}$ \ \ ;\ \ 
        $\mu_{q(1/\sigeps^2)}\leftarrow\smhalf(n+1)/B_{q(\sigeps^2)}$ 

\end{itemize}
until data no longer available or analysis terminated.
\end{itemize}
\vskip1mm
\hrule
\end{minipage}
\end{center}
\caption{\it Mean field variational Bayes algorithm for the determination
of the optimal parameters in $q^*(\bbeta,\bv)$, $q^*(\bgamma)$, 
$q^*(\sigma^2_u)$ and $q^*(\sigeps^2)$ for the Bayesian sparse signal
regression model (\ref{eq:penWavBayeIG}).}
\label{alg:sparseOnlMFVB} 
\end{algorithm}

As with the spline-based semiparametric regression models described in 
Section \ref{sec:LMM}, the wavelet-based models described here
benefit from the \emph{low-rank} property laid out in 
Section 3.1 of Wand \myand Ormerod (2011). This property entails
that the basis functions are fixed once and for all 
during the warm-up period. This permits fast updating of 
wavelet nonparametric fits as new data arrive. A cost of 
this approach is that the domain of predictors needs to be
specified based on the warm-up data. As explained in 
Section \ref{sec:LMM}, this will often
be reasonable. Of course, there is always the 
possibility of new predictor values landing outside 
domain of the basis functions, in which case some 
modification may be necessary.

Figure \ref{fig:waveOMFVB} illustrates online wavelet
nonparametric regression for data generated to 
$$\xnew\sim\mbox{Uniform(0,1)},\quad
\ynew|\,\xnew\sim N(\fWO(\xnew),1)$$
where $\fWO$ is defined by (20) of Wand \myand Ormerod (2011).
The warm-up sample size is $\nwarm=300$. The desired
improvement in the estimate of $\fWO$ as $n$ increases 
is clearly apparent. Convergence to the batch MFVB
estimate was found to be excellent in this case.

\ifthenelse{\boolean{ShowFigures}}
{
\begin{figure}[!ht]
\centering
{\includegraphics[width=0.85\textwidth]{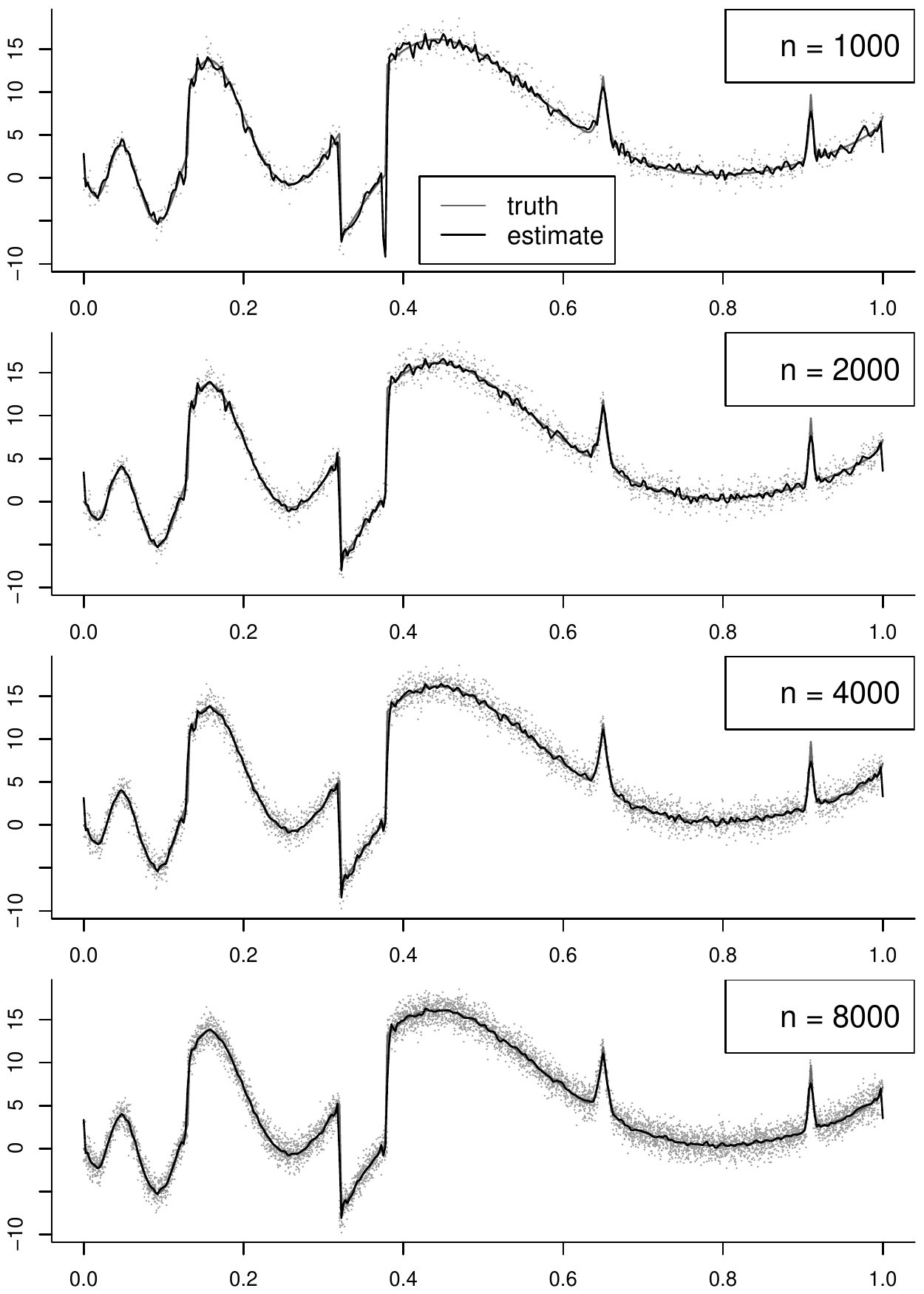}}
\caption{\it Examples of online MFVB wavelet fits
based on Algorithm \ref{alg:sparseOnlMFVB}. The true regression curve
is the function $\fWO$ defined in Wand \myand Ormerod (2011).}
\label{fig:waveOMFVB} 
\end{figure}
}
{\vskip3mm
\thickboxit{\bf \centerline{Wavelet illustration figure here.}}
\vskip3mm
}

\section{Binary Response Models}\label{sec:binResp}

The binary response model we consider here takes the same
form as (\ref{eq:LMM1}) and (\ref{eq:LMM2}), but with 
$\sigeps$ removed and
\begin{equation}
\by|\,\bbeta,\bu\sim\mbox{Bernoulli}\{\logit^{-1}(\bX\,\bbeta+\bZ\,\bu)\}.
\label{eq:logistMixMod}
\end{equation}
Note that (\ref{eq:logistMixMod}) is a convenient shorthand for the entries
of $\by$, conditional on $(\bbeta,\bu)$, being independent and with
$i$th entry $\mbox{Bernoulli}[\logit^{-1}\{(\bX\,\bbeta+\bZ\,\bu)_i\}]$.

Batch MFVB algorithms for approximate inference in (\ref{eq:logistMixMod}),
(\ref{eq:LMM1}) and (\ref{eq:LMM2}) start with the product restriction
$$p(\bbeta,\bu,a_{u1},\ldots,a_{ur},
\sigma_{u1}^2,\ldots,\sigma_{ur}^2)\approx
q(\bbeta,\bu,a_{u1},\ldots,a_{ur})\,   
q(\sigma_{u1}^2,\ldots,\sigma_{ur}^2).
$$
The resultant updates for the $\sigma_{u\ell}^2$ and $a_{u\ell}$ are
the same as in the Gaussian response case. The optimal $q$-density
for $(\bbeta,\bu)$ satisfies
\begin{equation}
q^*(\bbeta,\bu)\propto\exp\left\{
\by^T(\bX\bbeta+\bZ\bu)-\bone^T\log(1+e^{\bX\bbeta+\bZ\bu})
-\frac{1}{2\sigsqbeta}\Vert\bbeta\Vert^2-\smhalf
\sum_{\ell=1}^L\mu_{q(1/\sigma_{u\ell}^2)}\,\Vert\bu_{\ell}\Vert^2
\right\}.
\label{eq:qStarPreJJ}
\end{equation}
However, this is a non-standard form and poses tractability problems
with regards to approximate inference for $(\bbeta,\bu)$. A reasonable
remedy is to replace (\ref{eq:qStarPreJJ}) by a member of the 
following family of Multivariate Normal approximations:
$$\qunder^*(\bbeta,\bu)\sim N(\bmu_{\qunder(\bbeta,\bu;\bxi)},
                              \bSigma_{\qunder(\bbeta,\bu;\bxi)})
$$
where
$$\bSigma_{\qunder(\bbeta,\bu)}\equiv
\left[2\bC^T\mbox{diag}\{\lambda(\bxi)\}\,\bC+
\mbox{blockdiag}\{\sigma_{\beta}^{-2}\,\bI_p,
\mu_{q(1/\sigma_{u1}^2)}\bI_{K_1},\ldots,
\mu_{q(1/\sigma_{ur}^2)}\bI_{K_r}\}
\right]^{-1},
$$
$\bxi$ is an $n\times1$ vector of positive \emph{variational}
parameters, $\lambda(x)\equiv \tanh(x/2)/(4\,x)$, and
$$\bmu_{\qunder(\bbeta,\bu)}\equiv
\bSigma_{\qunder(\bbeta,\bu)}\,
\bC^T(\by-\smhalf\bone)
$$
with $\bC=[\bX\ \bZ]$ as before. This family of approximations 
is due to Jaakkola \myand Jordan (2000) and its genesis is 
given there. Section 3.1 of Ormerod \myand Wand (2010) explains 
this approximation strategy using notation similar 
to that used here. Jaakkola \myand Jordan (2000) also present
an Expectation-Maximization argument that results in
$$\bxi\leftarrow\sqrt{\mbox{diagonal}[
\bC\,\{\bSigma_{\qunder(\bbeta,\bu\,;\,\bxi))}+
\bmu_{\qunder(\bbeta,\bu;\bxi)}\,
\bmu_{\qunder(\bbeta,\bu\,;\,\bxi)}^T\}\,\bC^T]}
$$
being the optimal update for the $\bxi$ vector. 
Algorithm \ref{alg:logistOnlineMFVB} is the online 
MFVB algorithm that arises from appropriately
modifying the batch MFVB algorithm for 
(\ref{eq:logistMixMod}) with the Jaakkola \myand Jordan (2000)
strategy.

An alternative route to an online MFVB algorithm for 
binary response linear mixed models involves the 
probit link and the  Albert \myand Chib (1993) auxiliary
variable strategy. Batch MFVB algorithms for models of
this general type have been developed by 
Girolami \myand Rogers (2006) and  Consonni \myand Marin (2007).
Modification of these algorithms for the probit link
version of (\ref{eq:logistMixMod}) should lead to an 
algorithm that performs online approximate inference 
similar to that performed by Algorithm \ref{alg:logistOnlineMFVB}.

\begin{algorithm}
\begin{center}
\begin{minipage}[t]{151mm}
\hrule
\begin{itemize}
\item[1.] Perform batch-based tuning runs analogous
to those described in Algorithm 2'
and determine a warm-up sample size $\nwarm$ for which
convergence is validated.
\item[2.] Set $\bywarm$ and $\bCwarm$ 
to be the response vector
and design matrix, and  $\bxiwarm$ to be the vector
of variational parameters,
based on the first $\nwarm$ observations. 
Then set
$\bC^T\,(\by-\smhalf\bone)\leftarrow\bCwarm^T(\bywarm-\smhalf\bone)$,
\ $\bC^T\mbox{diag}\{\lambda(\bxi)\}\bC\leftarrow
\bCwarm^T\mbox{diag}\{\lambda(\bxiwarm)\}\bCwarm$,
\ $n\leftarrow\nwarm$. 
Also, set
$\bmu_{\qunder(\bbeta,\bu;\bxi)},
\ \bSigma_{\qunder(\bbeta,\bu;\bxi)},
\ \mu_{q(1/\sigma^2_{u1})},\ldots,\mu_{q(1/\sigma^2_{ur})}$
to be the values for these quantities obtained in the batch-based 
tuning run with sample size $\nwarm$.

\item[3.] Cycle:
\begin{itemize}
\item[] read in $\ynew\ (1\times 1)$ and $\bcnew\ (P\times 1)$
\ \ ;\ \ $n\leftarrow n+1$\\
\item[]$\xi\leftarrow\sqrt{
\bcnew^T\{\bSigma_{\qunder(\bbeta,\bu;\bxi)}+
\bmu_{\qunder(\bbeta,\bu;\bxi)}
\bmu_{\qunder(\bbeta,\bu;\bxi)}^T\}\bcnew}$
\item[] 
$\bC^T(\by-\smhalf\bone)\leftarrow\bC^T
(\by-\smhalf\bone)+\bcnew\,(\ynew-\smhalf)$
\item[]
$\bC^T\mbox{diag}\{\lambda(\bxi)\}\bC
\leftarrow\bC^T\mbox{diag}\{\lambda(\bxi)\}\bC+
\lambda(\xi)\,\bcnew\,\bcnew^T$
\item[] $\bSigma_{\qunder(\bbeta,\bu)}\leftarrow
\left[2\bC^T\mbox{diag}\{\lambda(\bxi)\}\,\bC+
\mbox{blockdiag}\{\sigma_{\beta}^{-2}\,\bI_p,
\mu_{q(1/\sigma_{u1}^2)}\bI_{K_1},\ldots,
\mu_{q(1/\sigma_{ur}^2)}\bI_{K_r}\}
\right]^{-1}$
\item[] $\bmu_{\qunder(\bbeta,\bu)}\leftarrow 
\bSigma_{\qunder(\bbeta,\bu)}\,
\bC^T(\by-\smhalf\bone)$
\item[] For $\ell=1,\ldots,r$\,:
\begin{itemize}
\item[]
$\mu_{q(1/a_{u\ell})}\leftarrow 
1/\{\mu_{q(1/\sigma_{u\ell}^2)}+A_{u\ell}^{-2}\}  $
\item[]
$\mu_{q(1/\sigma_{u\ell}^2)}\leftarrow 
\displaystyle{\frac{K_{\ell}+1}{2\,\mu_{q(1/\auell)}
+\Vert\bmu_{q(\bu_{\ell})}\Vert^2
+\mbox{tr}(\bSigma_{q(\bu_{\ell})})}}$
\end{itemize}
\end{itemize}
\item[] until data no longer available or analysis terminated.
\end{itemize}
\hrule
\end{minipage}
\end{center}
\caption{\it Online mean field variational Bayes algorithm for
approximate inference in the binary response logistic mixed model
(\ref{eq:logistMixMod}).}
\label{alg:logistOnlineMFVB} 
\end{algorithm}

Figure \ref{fig:binRespConvDiag} performs batch-based
convergence diagnostics for a binary response nonparametric
regression example. This is a special case of (\ref{eq:logistMixMod})
with $r=1$ and $\bZ$ containing spline basis functions.
New predictor/response pairs $(\xnew,\ynew)$ were generated
according to 
\begin{equation}
\xnew\sim\mbox{Uniform}(0,1),\quad
\ynew|\xnew\sim\mbox{Bernoulli}(\logit^{-1}(\cos(4\,\pi\,\xnew)
+2\,\xnew-1)).
\label{eq:binRespGen}
\end{equation}
%

\ifthenelse{\boolean{ShowFigures}}
{
\begin{figure}[!ht]
\centering
{\includegraphics[width=\textwidth]{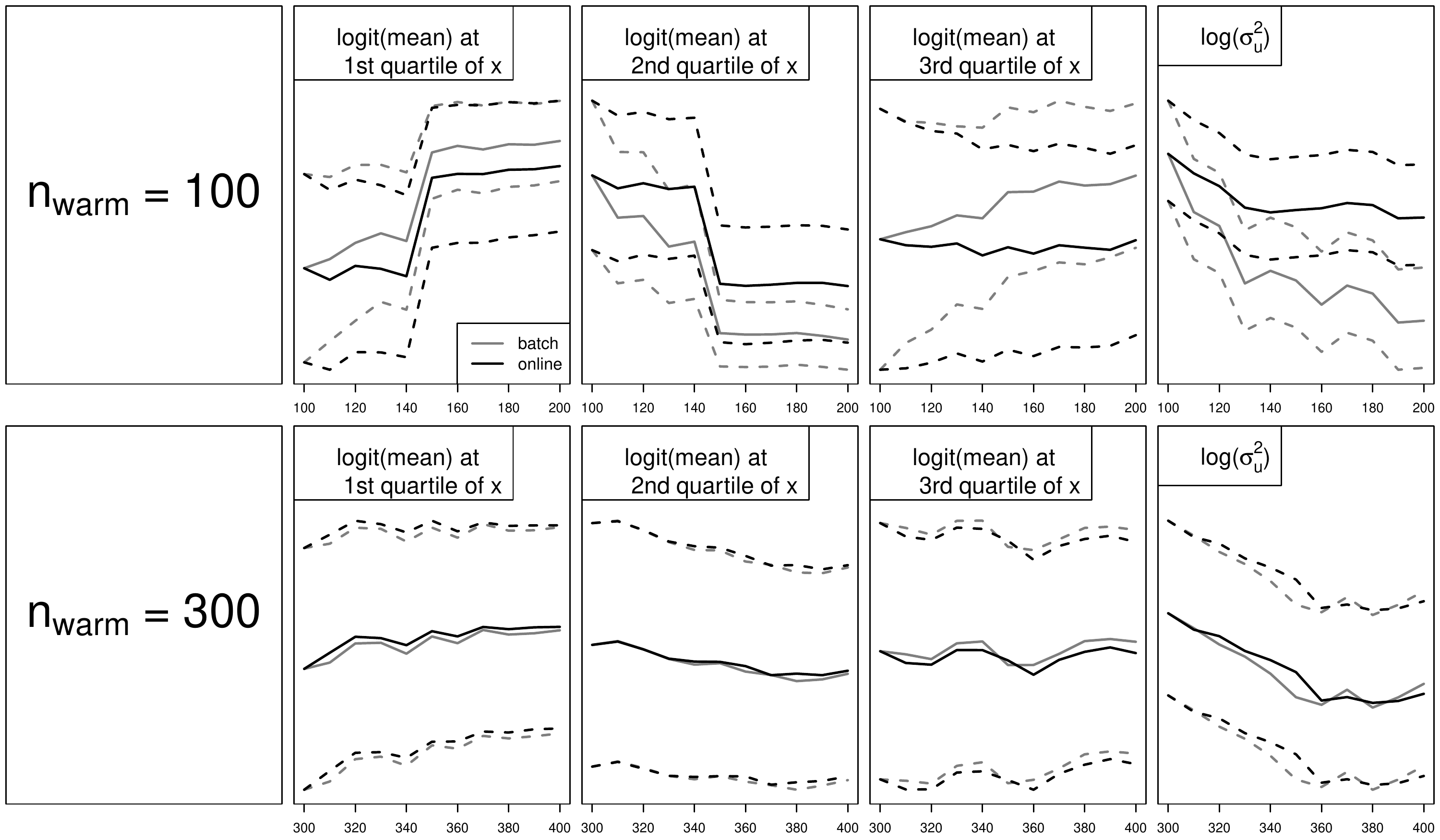}}
\caption{\it Convergence diagnostics for a binary 
response nonparametric regression example with data generated
according to (\ref{eq:binRespGen}).
The solid lines track the posterior means, whilst
the dashed lines show corresponding 95\% credible sets.
First row: the horizontal axes show the sample sizes between a warm-up batch
sample of size $\nwarm=100$ and validation sample sizes up to 
$\nvalid=100$ greater than $\nwarm$.  
Second row: as for the first row, but with $\nwarm=300$.}
\label{fig:binRespConvDiag} 
\end{figure}
}
{\vskip3mm
\thickboxit{\bf \centerline{Binary response diagnostic figure here.}}
\vskip3mm
}

\noindent

The analogues of Steps 1.-5. of Algorithm 2' were applied
with an initial trial involving $\nwarm=100$ and $\nvalid=100$.
The Bayes estimates and 95\% credible sets of the logit-transformed
mean function at each of the quartiles of the $x$-values, 
as well as $\log(\sigma_u^2)$, are shown in the upper row
of Figure \ref{fig:binRespConvDiag}. However, they have noticeable
disagreement, which indicates non-convergence of the online
MFVB results to their batch counterparts and that $\nwarm$ 
should be increased. Setting $\nwarm=300$ leads to the more 
concordant results shown in the lower row of Figure 
\ref{fig:binRespConvDiag}, indicating adequacy of this
warm-up size. We have found this behaviour typical for
binary response online MFVB and this simple example 
demonstrates the importance of batch-based tuning and convergence
diagnostics.

\section{Justification for Using Mean Field Variational Bayes}
\label{sec:justif}

Our use of online mean field variational Bayes 
is founded upon it being the only approach of which we are 
aware that (a) is readily extendible to a wide range of 
semiparametric regression models and (b), in 
the case of streaming data, has the ability to perform 
fast approximate inference for all model parameters.

Various other approaches such as stochastic gradient descent,
Markov chain Monte Carlo and expectation-maximisation  
can be ruled out since they fall short on at least one of 
these criteria. We now provide brief reasoning for 
their elimination from contention for real-time
semiparametric regression.

Stochastic gradient descent (e.g. Zhang, 2004) allows for 
regularized regression models to be fitted in an online
fashion. Recently Langford, Li \myand Zhang (2009) devised
stochastic gradient methodology for sparse signal regression.
However, in both Zhang (2004) and Langford {\it et al.} (2009),
the regularization parameters need to be inputted. This is
in contrast to Algorithms \ref{alg:lmmOnlineMFVB} and
\ref{alg:sparseOnlMFVB} in which the regularization parameters
are embedded in the underlying Bayesian model in the form
of variance parameters. This allows online estimation of the 
optimal amount of regularization. It appears that current
stochastic gradient descent technology does not support 
online estimation of regularization parameters.

Markov chain Monte Carlo (MCMC) has analogues with MFVB but
is much more computationally expensive. The full conditional 
distributions depend on the same matrix algebraic forms, such 
as $\by^T\by$, $\bC^T\by$ and $\bC^T\bC$, 
that appear in the batch MFVB algorithms for our 
semiparametric regression models. As shown in 
Algorithms \ref{alg:lmmOnlineMFVB}--\ref{alg:logistOnlineMFVB}, 
these forms are simple to update whenever a new vector 
of observations arrives. But MCMC then requires 
multiple sampling from the resulting full conditional
distributions. This is much more expensive than 
MFVB's arithmetic updates. For streaming data, 
this heavy computational burden will tend
to rule out MCMC.

Expectation-Maximization (EM) analogues of Algorithm 
\ref{alg:lmmOnlineMFVB}, but for frequentist linear
mixed models, are given in Sections 14.2a and 
14.2b of McCulloch, Searle \& Neuhaus (2008). They
are similar in nature to batch MFVB algorithms such
as Algorithm 3 of Ormerod \myand Wand (2010) and,
therefore, can be readily adapted for online processing.
Estimates of the precision are not included and further computing, 
possibly involving the Louis (1982) methodology, is
required for online inference. Moreover, the handling
of sparse shrinkage penalties and binary response variables 
requires considerably more complicated EM algorithms, 
and require approximation, such as Laplace's method, to 
be computationally feasible. In summary, an EM approach 
may lead to viable real-time semiparametric regression
algorithms, but they would be much more complicated
than Algorithms \ref{alg:lmOnlineMFVB}--\ref{alg:logistOnlineMFVB}.

Lastly, we mention Newton-Raphson optimization of 
the likelihood within a frequentist framework 
(e.g., Section 14.2c of McCulloch, 
Searle \& Neuhaus, 2008). For streaming data, there is
the problem of how to keep track of convergence of the 
Newton-Raphson schemes as data continually arise. 
The modification for sparse signal penalties looks particularly 
challenging. The binary response case also involves intractable 
forms which necessitate approximations such as those based on 
Laplace's method.

\section{Inferential Accuracy}\label{sec:infAccur}

Algorithms \ref{alg:lmOnlineMFVB}--\ref{alg:logistOnlineMFVB} 
perform real-time approximate Bayesian inference for the model
parameters. We now discuss the quality of the approximations
induced by the mean field assumptions.

Inferential accuracy of MFVB is a relatively new and modestly
studied area of statistical research. There have been a few
theoretical contributions, such as Wang \myand Titterington (2005),
and simulation studies, such as those presented in Faes, Ormerod 
\myand Wand (2011), but considerably more research is needed.
For the semiparametric regression models considered in the
present article, a broad summary is that MFVB exhibits good
to excellent inferential accuracy for the Gaussian response models
of Section \ref{sec:GaussResp} but only moderate to good accuracy for the
binary response model (\ref{eq:logistMixMod}). In particular, the 
approximate posterior density functions produced by Algorithm 
\ref{alg:logistOnlineMFVB} exhibit good accuracy for the variance
parameters. But for the coefficient vectors $\bbeta$ and $\bu$ the
approximate posterior density functions, whilst exhibiting good
locational behaviour, tend to under-approximate the spread.

Recently, Menictas \myand Wand (2013) provided some heuristic arguments,
based on likelihood theory, for why mean field approximations such as
(\ref{eq:prodRestrict}) and (\ref{eq:prodResLMM}) can be highly
accurate for Gaussian response models of Section \ref{sec:GaussResp}. 
The essential reason is parameter orthogonality between the 
coefficient parameters and variance parameters. 

Improving the accuracy of MFVB-based inference, especially 
for non-Gaussian response models such as Algorithm 
\ref{alg:logistOnlineMFVB}, is an important problem for
future research. For streaming data, a possible 
approach is to obtain batch MCMC-based fits in the warm-up phase 
and/or on parallel processors. These more accurate fits could
then be used to make appropriate corrections to the online
MFVB-based output. However, the details and efficacy of such
an approach are yet to be explored.

\section{Live Internet Demonstrations}\label{sec:illus}

We have launched the web-site: \texttt{realtime-semiparametric-regression.net}
for displaying live real-time semiparametric regression analyses. 
Links on this web-site point to several examples, and we anticipate
that the set of examples will grow during the next few years.
At the time of this writing, the examples involve simulated
data and three types of real-time data:  stock prices from the U.S.
\company{National Association of Securities Dealers Automated Quotations
(NASDAQ)} and the \company{London Stock Exchange} in the United Kingdom,
features of property rentals in Sydney, Australia, and data on 
delays in U.S. domestic flights.

\subsection{Simulated Data}

Our lead-off examples involve synthetic data.
First consider the Gaussian additive model
\begin{equation}
\begin{array}{l}
y_i|\bbeta,\bu_4,\bu_5,\bu_6,\sigeps^2\sim 
N\Big(\beta_1\,x_{1i}+\beta_2\,x_{2i}+\beta_3\,x_{3i}\\
\qquad\qquad +f_4(x_{4i})+f_5(x_{5i})+f_6(x_{6i}),\sigeps^2\Big)
\end{array}
\label{eq:GaussAddModel}
\end{equation}
where, for $j=4,5,6$, $\bu_j$ is vector of spline coefficients
for $f_j$. We generated 30,000 observations from (\ref{eq:GaussAddModel})
with $x_{1i},x_{2i},x_{3i}\simind\mbox{Bernoulli}\,(\smhalf)$
and $x_{4i},x_{5i},x_{6i}\simind N(0,1)$. Truth was set according
to $\beta_1=0.2$, $\beta_2=-0.3$, $\beta_3=0.6$, 
$f_4(x)=2\Phi(6x-3)$, $f_5(x)=\sin(3\pi x^3)$, 
$f_6(x)=\cos(4\pi x)$ and $\sigeps^2=1$. 
The link \texttt{Gaussian additive model} on the abovementioned
web-site points
to a movie showing summaries of the regression fits when the
data are sequentially fed into Algorithm \ref{alg:lmmOnlineMFVB}.

The \texttt{Logistic additive model} link points to a similar movie,
but with data generated from the logistic additive model
$$
y_i\,|\,\bbeta,\bu_2,\bu_3\sim 
\mbox{Bernoulli}\big(
\logit^{-1}(\beta_1\,x_{1i}+f_2(x_{2i})
+f_3(x_{3i}))\big)
$$
with $x_{1i}\simind\mbox{Bernoulli}\,(\smhalf)$,
$x_{2i},x_{3i}\simind N(0,1)$ and truth set at
$\beta_1=0.2$, $f_2(x)=\cos(4\pi\,x)+2\,x$ and
$f_3(x)=\sin(2\pi\,x^2)$. 

Lastly, the \texttt{Wavelet regression} link corresponds to the 
simulation setting used to produce Figure \ref{fig:waveOMFVB},
with description given in Section \ref{sec:sparse}.

\subsection{Stock Price Data}

In this set of examples, the predictor and response 
variable pairs correspond to pairs of stock prices. An example 
nonparametric regression model is
\begin{equation}
(\mbox{\texttt{Microsoft\,\,stock\,\,price}})_i|\,\bbeta,\bu,
\sigeps^2\simind 
N(\beta_0+f(\mbox{\texttt{(Intel\,\,stock\,\,price)}}_i),\sigeps^2)
\label{eq:NASDAQ}
\end{equation}
where $f(x)=\beta_1\,x+\sum_{k=1}^K\,u_k\,z_k(x)$ is a penalized
spline function as described in Section \ref{sec:LMM}
with the same distributional structures imposed on the model
parameters. In addition,
$(\mbox{\texttt{Microsoft stock price}})_i$ and 
$(\mbox{\texttt{Intel}}$ $\mbox{\texttt{stock price}})_i$ 
denote the $i$th stock price for the U.S. companies 
\company{Microsoft Corporation} and 
\company{Intel Corporation}, respectively, for the current trading day.
The web-site displays fitting
of (\ref{eq:NASDAQ}) in real-time during the 
\company{NASDAQ} opening hours (9:30am to 4:00pm North American
Eastern Standard Time). The \textsf{R} package \texttt{quantmod}
(Ryan, 2012) is used to obtain the \company{NASDAQ} data from the 
\company{Yahoo!\ Finance} web-site (\texttt{finance.yahoo.com}). 

A similar series of examples is set up using \company{London Stock Exchange}
data during stock market opening hours 
(8:00 am to 4:20 pm Greenwich Mean Time). Note that 
\company{Yahoo!\ Finance} delays \company{London Stock Exchange}
data by 20 minutes.

Depending on the example and the live data-set, 
the appropriateness of the nonparametric
regression model (\ref{eq:NASDAQ}) may be questionable
and more sophisticated models could be entertained.
Hence, these examples should only be viewed as simple
illustrations of the concept of real-time semiparametric
regression.

\subsection{Sydney Property Rental Data}

This example involves real-time semiparametric regression 
analysis of data from the property rental market in Sydney, Australia. 
Each day, hundreds of properties come on the Sydney market and these fresh 
data are usually advertised on rental agency web-sites and real estate 
web-sites as \texttt{realestate.com.au}. This offers the possibility 
to perform real-time analysis and produce live and up-to-date summaries of 
the rental market status. An attractive approach to model such data is the 
special case of semiparametric regression known as 
geoadditive models (Kammann and Wand, 2003). Explicitly, we work
with the model
\begin{equation}
\begin{array}{l}
\log((\mbox{\texttt{weekly\,\,rent}})_{ij})|\,\bbeta,U_i,
\bu_2,\bu_3,\bu_4,\bu_5,
\sigeps^2\simind\\[1ex] 
\qquad
N(\beta_0 + \beta_1\,\mbox{\texttt{house}}_{ij} +
f_2(\mbox{\texttt{(number\,\,of\,\,bedrooms)}}_{ij})\\[1ex]
\qquad + f_3(\mbox{\texttt{(number\,\,of\,\,bathrooms)}}_{ij} )
+f_4(\mbox{\texttt{(number\,\,of\,\,car\,\,spaces)}}_{ij} )\\[1ex] 
\qquad+ f_5(\mbox{\texttt{longitude}}_{ij},\mbox{\texttt{latitude}}_{ij})
+U_i,\sigeps^2),
\null\qquad U_1,\ldots,U_{992}|\,\sigma_U^2\simind N(0,\sigma_U^2).
\end{array}
\label{eq:SydneyRealEstate}
\end{equation}
Here, $\mbox{(\texttt{weekly rent})}_{ij}$ is 
the weekly rental amount in Australian dollars of the $j$th property
for the $i$th real estate agency (hereafter called the $(i,j)$th property), 
and $\mbox{\texttt{house}}_{ij}$ is an indicator of the $(i,j)$th property 
being a house, townhouse or villa (rather than an apartment).
The variable $\mbox{\texttt{(number\,\,of\,\,bedrooms)}}_{ij}$ 
is the number of bedrooms in the $(i,j)$th property.
Variables concerning the numbers of bathrooms and car spaces are
defined similarly. The geographical location of the $(i,j)$th property
is conveyed by the variables $\texttt{longitude}_{ij}$
and $\texttt{latitude}_{ij}$. The  $U_i$, $1\le i\le 992$, 
are random intercepts for each of the 992 agencies. The fixed effect regression
coefficients $\beta_0$, $\beta_1$  and the linear contribution
to $f_2,\ldots,f_5$ are stored in $\bbeta$. 
Similarly, the spline basis coefficients for $f_2,\ldots,f_5$
are stored in $\bu_2,\ldots,\bu_5$. The estimate of $f_5$ is based
on bivariate thin plate splines as explained in Chapter 13
of Ruppert, Wand \myand Carroll (2003).

The web-site for this example displays fitting of 
(\ref{eq:SydneyRealEstate}) in real time based on
data collected since 9th May, 2012.
Several regression summaries are presented. 
Firstly, a geographical map is listed with processed properties as small black 
dots and recently (i.e. during the last hour) added ones as yellow 
circles. The total number of processed properties is included at the 
bottom right. Next, a color-coded geographical map displays the weekly 
rent for a two bedroom apartment with one bathroom and one car space 
for various geographical locations. The approximate posterior density function
for $\beta_1$ shows the impact of the property being a house or not.
Regression fits and 95\% credible sets for the number of bedrooms, 
bathrooms and car spaces for apartments are presented. 
Finally, a list of rental agencies with the least and most 
expensive properties, after correcting 
for all other covariates, is provided. All these regression summaries are 
computed in real time and the figures are updated every hour.

\subsection{U.S. Domestic Flight Data}

Air traffic delays represent a critical problem for both airlines and 
passengers. In this section we will demonstrate the proposed 
methodology for real-time analysis of U.S. domestic flights.  
We use the web-site \texttt{www.flightstats.com} 
to obtain real-time data on flight delay, flight distance,
operating airline and flight path. Data on temperature, 
wind speed and aviation flight category is obtained through the
\texttt{aviationweather.gov} web-site. This example is inspired
by a recent competition, titled \company{GE Flight Quest}, run
by the \company{kaggle} platform (\texttt{www.kaggle.com}).

The real-time data consist of flight delay, flight distance, 
operating airline and flight path. In addition, data on temperature, 
wind speed and aviation flight category are available
The  aviation flight categories are based on the North American
conventions known as METAR and are based on
the ceiling (height above
ground of the base of the lowest layer of cloud) and visibility.
Table \ref{tab:METAR} provides the aviation flight categories definitions.

\begin{table}[ht]
\begin{center}
\begin{tabular}{lll}
category                     & ceiling    &  and/or visibility \\[0.5ex]
\hline
visual flight rules          & above 3,000 feet & above 5 miles\\[0.5ex]
marginal visual flight rules & 1000--3,000 feet & 3--5 miles \\[0.5ex]
instrument flight rules      & 500--1,000  feet & 1--3 miles\\[0.5ex] 
low instrument flight rules  & below 500 feet  & below 1 mile\\[0.5ex]
\hline
\end{tabular}
\end{center}
\caption{\it Definitions of North American aviation flight categories.}
\label{tab:METAR} 
\end{table}

Our demonstration uses the semiparametric regression model:
\begin{equation}
\begin{array}{l}
\log(\mbox{\texttt{delay}}_{ijk}+120)|\,\bbeta,U_i, V_j,\bu_7,
\bu_8,\bu_9,\bu_{10},\bu_{11},\sigeps^2\simind \\[1ex]
\hspace{2mm} N(\beta_0 + \beta_1 {\mbox{\texttt{MVFRdep}}}_{ijk} 
+ \beta_2 {\mbox{\texttt{IFRdep}}}_{ijk} + \beta_3 {\mbox{\texttt{LIFRdep}}}_{ijk}  
+ \beta_4 {\mbox{\texttt{MVFRarr}}}_{ijk}\\[1ex]
\hspace{2mm} + \beta_5 {\mbox{\texttt{IFRarr}}}_{ijk} 
+ \beta_6 {\mbox{\texttt{LIFRarr}}}_{ijk}  + f_7({\mbox{\texttt{(flight distance)}}}_{j} )
\\[1ex]
\hspace{2mm} + f_8({\mbox{\texttt{(departure temperature)}}}_{ijk} ) 
+ f_9({\mbox{\texttt{(arrival temperature)}}}_{ijk} )
\null\\[1ex]
\hspace{2mm} +  f_{10}({\mbox{\texttt{(departure wind speed)}}}_{ijk} ) 
+ f_{11}({\mbox{\texttt{(arrival wind speed)}}}_{ijk} )
\null\\[1ex]
\hspace{2mm} + U_i + V_j,\sigeps^2),
\hspace{3mm} U_1,\hdots,U_{171} |\,\sigma^2_{U} \simind N(0,\sigma^2_{U}),
\hspace{3mm} V_1,\hdots,V_{2,000} |\,\sigma^2_{V} \simind N(0,\sigma^2_{V}).
\end{array}
\label{eq:DomesticFlightData}
\end{equation}

Here \texttt{delay}$_{ijk}$ is the difference between the 
actual and scheduled runway arrival time in minutes for the $k$th flight 
of airline $i$ on flight path $j$ and 
$$
\texttt{MVFRdep}_{ijk}=\left\{ 
\begin{array}{ll}
1 & \mbox{if marginal visual flight rules apply at the scheduled runway }\\ 
  & \mbox{departure time of the $k$th flight of airline $i$ on flight path $j$}\\ 
0 & \mbox{otherwise.}
\end{array}
\right.
$$
The variable $\texttt{MVFRarr}_{ijk}$ is defined analogously, but
for the scheduled runway arrival time.
The other aviation flight category variables are defined
similarly, with \texttt{IFR} denoting ``instrument flight rules''
and \texttt{LIFR} denoting ``low instrument flight rules''.
The variable \texttt{(flight} \texttt{distance)}$_{j}$ 
denotes the distance of flight path $j$ in kilometers. 
Variables \texttt{(departure temperature)}$_{ijk}$ and 
\texttt{(arrival temperature)}$_{ijk}$ are the temperature in 
degrees Celsius at the scheduled runway departure and arrival 
time of the $k$th flight of airline $i$ on flight path $j$, 
respectively. Variables \texttt{(departure wind speed)}$_{ijk}$ 
and \texttt{(arrival wind speed)}$_{ijk}$ are the wind speed in 
knots at the scheduled runway departure and arrival time of the 
$k$th flight of airline $i$ on flight path $j$, respectively. 
The $U_i$,$ 1 \leq i \leq 171$, are random intercepts for each 
of the 171 airlines, while $V_j$,$ 1\leq j \leq$ 2,000, are random 
effects for each
of the 2,000 flight paths. The fixed effect regression coefficients 
$\beta_0,\hdots,\beta_6$ and the linear contribution to 
$f_7, \hdots , f_{11}$ are stored in $\bbeta$. Similarly, the 
spline basis coefficients for $f_7, \hdots , f_{11}$ are stored 
in $\bu_7, \hdots ,\bu_{11}$.\\

The link \texttt{U.S. domestic flight data} on our
live demonstrations web-site displays fitting of 
(\ref{eq:DomesticFlightData}) in real time based on data collected since 
25th January, 2013. A map shows the flight paths that have most 
recently been processed and the number of processed flights 
is given at the bottom of the map. Various regression summaries are provided.
Of particular interest are tables of airlines and flight paths with
the lowest and highest delays. All these regression summaries are 
computed in real-time and the figures are updated every few minutes.

\section*{Acknowledgments}

This research was partially supported
by Australian Research Council Discovery Project
DP110100061. T. Broderick's research was supported
by a U.S. National Science Foundation Graduate Research Fellowship.
We are grateful to Jeff Morris and Paul Murrell 
for discussions related to this research.

\setstretch{1.0}
\section*{References}

\bib
Albert, J.H. \myand Chib, S. (1993). Bayesian analysis of
binary and polychotomous response data.  
{\it Journal of the American Statistical Association}, 
{\bf 88}, 669--679.

\bib
Armagan, A., Dunson, D.B. \myand Lee, J. (2012).
Generalized double Pareto shrinkage.
{\it Statistica Sinica}, to appear.

\bib
Bishop, C.M. (2006). {\it Pattern Recognition and Machine Learning.}
New York: Springer.

\bib
Cameron, A.C. \myand Trivedi, P.K. (2005).
\textit{Microeconometrics: Methods and Applications.}
New York: Cambridge University Press.

\bib
Carvalho, C.M., Polson, N.G. \myand Scott, J.G. (2010).
The horseshoe estimator for sparse signals.
{\it Biometrika}, {\bf 97}, 465--480.

\bib
Consonni, G. \myand Marin, J.-M. (2007).
Mean-field variational approximate Bayesian inference for 
latent variable models. {\it Computational Statistics
and Data Analysis}, {\bf 52}, 790--798.

\bib
Croissant, Y. (2011). Ecdat 0.1.
Data sets for econometrics. R package,\\
\texttt{cran.r-project.org}

\bib
Devroye, L. \myand Wagner, T.J. (1980).
On the $L_1$ convergence of kernel estimators of
regression functions with application to discrimination.
\textit{Zeitschrift f\"ur Wahrscheinlichkeitstheorie
und Vervandte Gebiete}, {\bf 51}, 15--25.

\bib
Faes, C., Ormerod, J.T. \myand Wand, M.P. (2011).
Variational Bayesian inference for parametric
and nonparametric regression with missing data.
{\it Journal of the American Statistical Association},
{\bf 106}, 959--971.

\bib
Fricker, R.D. \myand Chang, J.T. (2008).
A spatio-temporal methodology for real-time
biosurveillance. \textit{Quality Engineering}, 
{\bf 20}, 465--477.

\bib
Girolami, M. \myand Rogers, S. (2006). Variational Bayesian 
multinomial probit regression. {\it Neural Computation}, 
{\bf 18}, 1790--1817.

\bib
Gradshteyn, I.S. \myand Ryzhik, I.M. (1994). 
{\it Tables of Integrals, Series, and Products}, 5th Edition,
San Diego, California: Academic Press. 

\bib
Griffin, J.E. \myand Brown, P.J. (2011).
Bayesian hyper lassos with non-convex penalization.
{\it Australian and New Zealand Journal of Statistics},
{\bf 53}, 423--442.

\bib
H\"ardle, W. (1990). \textit{Applied Nonparametric Regression.}
Cambridge: Cambridge University Press.

\bib
Hoffman, M., Blei, D. \myand Bach, F. (2010).
Online learning for latent Dirichlet allocation.
In  \textit{Advances in Neural Information Processing Systems 23},
Lafferty, J., Williams, C.K.I., Shawe-Taylor, J., 
Zemel, R.S. \myand Culotta, A. (eds.)
pp. 856--864.

\bib
Huang, A. \myand Wand, M.P. (2012).
Simple marginally noninformative prior distributions
for covariance matrices. Under revision for 
{\it Bayesian Analysis}.\\
\texttt{www.uow.edu/}$\sim\,$\texttt{mwand/papers.html}

\bib
Jaakkola, T.S. \myand Jordan, M.I. (2000). 
Bayesian parameter estimation via variational methods.
{\it Statistics and Computing} {\bf 10}, 25--37.

\bib
Jank, W. \myand Shmueli, G. (2007). 
Modelling concurrency of events in on-line auctions via
spatiotemporal semiparametric models.
{\it Applied Statistics}, {\bf 56}, {1--27}.

\bib
Johnstone, I.M. \myand Silverman, B.W. (2005).
Empirical Bayes selection of wavelet thresholds.
{\it The Annals of Statistics}, {\bf 33}, 1700--1752.

\bib
Kaimi, I. \myand Diggle, P.J. (2011).
A hierarchical model for real-time monitoring 
of variation in risk of non-specific gastro-intestinal infections.
\textit{Epidemiology and Infection}, {\bf 139}, 
1854--1862. 

\bib
Kammann, E.E. and Wand, M.P. (2003). Geoadditive models.
{\it Journal of the Royal Statistical Society, Series C}, {\bf 52}, 1--18.

\bib
Krzyzak, A. \myand Pawlak, M. (1984).
Almost everywhere convergence of a recursive 
regression function estimate and classification.
\textit{IEEE Transactions on Information
Theory}, {\bf IT-30}, 91--93.

\bib
Langford, J., Li, L. \myand Zhang, T. (2009).
Sparse online learning via truncated gradient.
\textit{Journal of Machine Learning Research},
{\bf 10}, 777--801.

\bib
Louis, T.A. (1982). Finding the observed information
matrix when using the EM algorithm.
{\it Journal of the Royal Statistical Society, Series B}, 44, 226--233.

\bib
Luenberger, D.G. \myand Ye, Y. (2008). 
\textit{Linear and Nonlinear Programming}. 
New York: Springer, 3rd edition.

\bib
McCulloch, C.E., Searle, S.R. \myand Neuhaus, J.M. (2008). 
{\it Generalized, Linear, and Mixed Models}, 2nd Edition.
John Wiley \& Sons, New York.

\bib
Menictas, M. \myand Wand, M.P. (2013).
Variational inference for marginal longitudinal 
semiparametric regression. \textit{Stat}, in press.

\bib
Michalak, S., DuBois, A., DuBois, D., Vander Wiel, S. 
\myand Hogden, J. (2012).
Developing systems for real-time streaming analysis.
\textit{Journal of Computational and Graphical 
Statistics}, {\bf 21}, 561--580.

\bib
Neville, S.E., Ormerod, J.T. \myand Wand, M.P. (2012).
Mean field variational Bayes for continuous sparse
signal shrinkage: pitfalls and remedies.\\
\texttt{www.uow.edu/}$\sim\,$\texttt{mwand/papers.html}

\bib
Ng, S-K., McLachlan, G.J. \myand Lee, A.H. (2006).
An incremental EM-based learning  approach for on-line prediction 
of hospital resource utilization. \textit{Artificial Intelligence in Medicine}, 
{\bf 36}, 257--267.

\bib
Ormerod, J.T. \myand Wand, M.P. (2010).
Explaining variational approximations.
{\it The American Statistician}, 
{\bf 64}, 140--153.

\bib
Ruppert, D., Wand, M.P. \myand Carroll, R.J. (2003). 
{\it Semiparametric Regression}.
New York: Cambridge University Press. 

\bib
Ruppert, D., Wand, M.P. \myand Carroll, R.J. (2009).
Semiparametric regression during 2003-2007.
{\it Electronic Journal of Statistics}, {\bf 3}, 1193--1256.

\bib
Ryan, J.A. (2012) quantmod 0.3.
Quantitative financial modelling
framework. R package,\\
\texttt{cran.r-project.org}

\bib
Smith, A.D.A.C. \myand Wand, M.P. (2008).
Streamlined variance calculations for semiparametric
mixed models. {\it Statistics in Medicine}, {\bf 27}, 435--448.

\bib
Tchumtchoua, S., Dunson, D.B. \myand Morris, J.S. (2012).
Online variational Bayes inference for high-dimensional
correlated data. Unpublished manuscript.\\
\texttt{www.stat.duke.edu/}$\sim\,$\texttt{dunson/submitted.html}

\bib
Wainwright, M.J. \myand Jordan, M.I. (2008).
Graphical models, exponential families, and variational inference.
{\it Foundation and Trends in Machine Learning}, {\bf 1}, 1--305.

\bib
Wand, M.P. (2009).
Semiparametric regression and graphical models.
{\it Australian and New Zealand Journal
of Statistics}, {\bf 51}, 9--41.

\bib
Wand, M.P. \myand Jones, M.C. (1995)
{\it Kernel Smoothing}. London: Chapman and Hall.

\bib
Wand, M.P. \myand Ormerod, J.T. (2008).
On semiparametric regression with O'Sullivan penalized splines.
{\it Australian and New Zealand Journal of Statistics},
{\bf 50}, 179--198.

\bib
Wand, M.P. \myand Ormerod, J.T. (2011).
Penalized wavelets: embedding wavelets
into semiparametric regression.
{\it Electronic Journal of Statistics},
{\bf 5}, 1654--1717.

\bib
Wang, C., Paisley, J. \myand Blei, D.M. (2011).
Online variational inference for the hierarchical
Dirichlet process. 
\textit{International Conference on 
Artificial Intelligence and Statistics, 2011},
Fort Lauderdale, Florida, USA.

\bib
Wang, B. \myand Titterington, D.M. (2005).
Inadequacy of interval estimates corresponding to variational
Bayesian approximations. In \textit{Proceedings of the 10th
International Workshop on Artificial Intelligence}, eds.
R.G. Cowell and Z. Ghahramani, Barbados: Society for
Artificial Intelligence and Statistics, pp. 373--380.

\bib
Welham, S.J., Cullis, B.R., Kenward, M.G. \myand Thompson, R. (2007). 
A comparison of mixed model splines for curve fitting.
{\it Australian and New Zealand Journal of Statistics},
{\bf 49}, 1--23.

\bib
Wolverton, C.T. \myand Wagner, T.J. (1969).
Asymptotically optimal discriminant functions for 
pattern recognition. \textit{IEEE Transactions on Information
Theory}, {\bf IT-15}, 258--265.

\bib
Wood, S.N. (2006).
{\it Generalized Additive Models: An Introduction with R.}
Boca Raton, Florida: Chapman \& Hall/CRC.

\bib
Yamato, H. (1971). Sequential estimation of a continuous probability
density function and model. \textit{Bulletin of Mathematical
Statistics}, {\bf 14}, 1--12.

\bib
Zhang, T. (2004). Solving large scale linear prediction problems
using stochastic gradient descent algorithms. In
\textit{Proceedings of the Twenty-First International Conference
on Machine Learning}, Brodley, C.E. (ed.). 
pp. 919--926. 

\bib
Zhao, Y., Staudenmayer, J., Coull, B.A. \myand Wand, M.P. (2006).
General design Bayesian generalized linear mixed models.
{\it Statistical Science}, {\bf 21}, 35--51.

\end{document}